\documentclass[11pt]{article}
\usepackage[utf8]{inputenc}
\usepackage{amsmath, mathrsfs, amsthm}
\usepackage{amssymb}
\usepackage{mathtools}

\usepackage{graphicx,psfrag,epsf}
\usepackage{enumerate}
\usepackage[export]{adjustbox}

\usepackage{subfig}

\newtheorem{theorem}{Theorem} 

\newtheorem{remark}{Remark}

\usepackage{natbib}

\usepackage{times}
\usepackage{bm}

\usepackage{amssymb}

\usepackage[plain,noend]{algorithm2e}

\usepackage{hyperref}

\renewenvironment{align}
{\begin{equation}\begin{aligned}}
{\end{aligned}\end{equation}}

\protected\def\[#1\]{\begin{equation}\begin{aligned}#1\end{aligned}\end{equation}}
\protected\def\(#1\){\begin{equation*}\begin{aligned}#1\end{aligned}\end{equation*}}

\usepackage{float}

\begin{document}

\title{Low Tree-Rank Bayesian Vector Autoregression Models}

\author{Leo L Duan
\thanks{Department of Statistics, University of Florida, {li.duan@ufl.edu}}
\quad\quad
Zeyu Yuwen
\thanks{Department of Statistics, University of Florida, {zeyu.yuwen@ufl.edu}}
 \quad\quad
George Michailidis
 \thanks{Department of Statistics, University of California at Los Angeles, {gmichail@g.ucla.edu}}
 \quad\quad
 Zhengwu Zhang
\thanks{
 Department of Statistics and Operations Research, University of North Carolina at Chapel Hill, {zhengwu\_zhang@unc.edu}}
}

\date{}
\maketitle

\begin{abstract}
Vector autoregression has been widely used for modeling and analysis of multivariate time series data. In high-dimensional settings, model parameter regularization schemes inducing sparsity yield interpretable models and achieved good forecasting performance. However, in many data applications, such as those in neuroscience, the Granger causality graph estimates from existing vector autoregression methods tend to be quite dense and difficult to interpret, unless one compromises on the goodness-of-fit.  To address this issue, this paper proposes to incorporate a commonly used structural assumption  --- that the ground-truth graph should be largely connected, in the sense that it should only contain at most a few components.  We take a Bayesian approach and develop a novel tree-rank prior distribution for the regression coefficients. Specifically, this prior distribution forces the non-zero coefficients to appear only on the union of a few spanning trees. Since each spanning tree connects $p$ nodes with only $(p-1)$ edges, it effectively achieves both high connectivity and high sparsity. We develop a computationally efficient Gibbs sampler that is scalable to large sample size and high dimension.
In analyzing test-retest functional magnetic resonance imaging data, our model produces a much more interpretable graph estimate, compared to popular existing approaches. In addition,  we show appealing properties of this new method, such as efficient computation, mild stability conditions and posterior consistency.
\vskip 12pt
{\noindent  KEYWORDS: }  Graph rank, Laplacian matrix,  Structural vector autoregression, Gibbs sampling, Neuroimaging data
\end{abstract}

\section{Introduction}
\label{sec:intro}
Vector autoregression (VAR) models have been widely used for modeling multivariate time series data in economics \citep{eichler2007granger,stock2016dynamic,lin2020regularized}, genomics \citep{michailidis2013autoregressive,basu2015network} and neuroscience \citep{Anil2015}. The observations $y^t=(y^t_1,\ldots, y^t_p)\in \mathbb{R}^p$ at discrete time points $t=1,\cdots, \mathcal T$ evolve according to:
\begin{align}\label{eq:var}
y^t= C^{(1)}  y^{t-1}+C^{(2)} y^{t-2}+...+C^{(d)} y^{t-d} +\varepsilon^t,
\end{align}
where the transition matrices $C^{(k)} \in \mathbb{R}^{p \times p}$ capture lead-lag effects at lags $1,\cdots,d$ and $\varepsilon^t \in \mathbb{R}^p$ is a noise term.
The elements of the transition matrices $C^{(k)}$ form a directed graph of Granger causal effects \citep{Granger1969}; specifically, if there is at least one $C^{(k)}_{i,j} \neq 0$ for some $k=1,\cdots,d$, then it implies that $y_j^{t}$ is predictive for future values of $y_i^{t'}$ with $t' \ge t+k$, and an edge is included $(j\to i)$ in the corresponding graph.

However, in many applications, the dimensionality of the parameter space $p^2d$ exceeds the number of available observations. To overcome this challenge, several Bayesian and frequentist regularization approaches have been proposed in the literature. For example, \cite{Sims1993} proposed to impose a Gaussian prior distribution on the elements of the transitions matrices, whereas \cite{Banbura2010} used a Gaussian-inverted Wishart prior distribution to induce ridge type shrinkage. \cite{Dimitris2013} put Bernoulli prior distributions on the indicators of each parameter in the transition matrix to select Granger causal effects. More recently, \cite{Ghosh2019,ghosh2021strong} studied the theoretical properties of Bayesian VAR models under various prior distributions for the parameters and
established their posterior consistency. In frequentist approaches, various sparsity-inducing penalties have been proposed and studied. \cite{Basu2015} used a lasso penalty and developed key technical results to establish estimation consistency of the model parameters. Variants of sparse regularization schemes were proposed in \cite{Anders2015,lin2017regularized,Hsu2008,Nicholson2020}. A different direction was pursued by \cite{Basu2019} that assumes that the transition matrices exhibit a low-rank and sparse structure. Another variant integrates  additional data summarized as factors that are incorporated as additional time series in the model \citep{lin2020regularized}.

These regularized versions of the vector autoregressive model generally exhibit very good predictive performance. However, in many cases, the resulting Granger causal graph is fairly dense, which makes interpretation more challenging, and/or \textit{disconnected}, which contradicts scientific background knowledge in certain application domains. Indeed, in the neuroimaging application discussed in Section \ref{sec:application}, existing sparsity-inducing approaches produce very dense Granger causal graphs unless the tuning parameters that control the degree of regularization are selected to produce much sparser estimates at the expense of a significantly poorer goodness-of-fit.

To address this challenge, we introduce a model that posits the Granger causal graph to be \textit{connected} (or almost connected) and containing relatively few edges, thus making it highly interpretable and suitable for applications wherein the underlying science dictates full connectivity. We achieve this by developing a novel \textit{tree-rank} prior distribution and the corresponding algorithm to calculate the posterior distribution of the model parameters, and establishing its theoretical properties. The proposed model has been employed to estimate robust Granger causal graphs from functional MRI (fMRI) data obtained from the Human Connectome Project. Granger causal graphs play an important role in fMRI analysis, primarily due to their ability to examine directional relationships, or causal influences, between different brain regions. 

In recent developments in the domain of network neuroscience, tree-type connectivity has received considerable attention. A review of neurophysiological and neuroimaging studies \citep{blomsma2022minimum} suggests that line-like tree organization characterizes neurodegenerative disorders across pathologies and is associated with symptom severity and disease progression. In an Alzheimer's disease (AD) study \citep{guo2017alzheimer}, it was reported that the minimum spanning tree extracted from high-order functional connectivity greatly improves the diagnostic accuracy for AD. In a dementia study \citep{saba2019brain}, it was found that brain connectivity, characterized by spanning tree estimates and the degree of possible breakdowns in information flow, is highly associated with the behavioral variants of frontotemporal dementia. These are just selected examples from a vast and rapidly developing neuroscience literature, suggesting that the assumption of tree connectivity structure is fairly plausible for the brain network. We have also demonstrated in our data application that incorporating such an assumption into the statistical model could significantly improve the accuracy and reproducibility of the graph estimate.

The remainder of the paper is organized as follows. In Section \ref{sec:model}, we introduce the tree-rank VAR model and develop a Gaussian scale mixture prior on the coefficient matrix. In Section \ref{sec:computation}, we introduce the posterior distribution initialization and computation. In Section \ref{sec:stability}, we establish a mild stability condition for the model, while in Section \ref{sec:consistency}, posterior consistency and model selection consistency. Sections \ref{sec:simulation} and \ref{sec:application} illustrate the performance of tree-rank estimates on synthetic and resting-state functional magnetic resonance imaging data. We conclude with a discussion in Section \ref{sec:discussion}. The software is available at \href{https://github.com/leoduan/Spanning-Tree-VAR}{https://github.com/leoduan/Spanning-Tree-VAR}.

\section{Modeling Framework}\label{sec:model}
\subsection{Vector Autoregressive Processes from a Tree-Covered Graph}\label{sec:VAR-tree}

The underlying data generating process corresponds to the VAR model in \eqref{eq:var}, with $\{C^{(k)}, k=1,\cdots,d\}$ transition matrices defining the Granger causal network $G=(V,E)$ \citep{basu2015network}. Specifically, if there is an edge $(j\to i) \in E$, then there is at least one $C^{(k)}_{i,j}\neq 0$ for $k=1,\cdots,d$. Further, we assume Gaussian measurement error $\varepsilon^t \stackrel{iid}{\sim}  \text{N}(0,\Sigma_\varepsilon)$ for all $t$, with some positive definite covariance $\Sigma_\varepsilon$. To facilitate computation, we impose a near low-rank structure on the error covariance matrix $\Sigma_\varepsilon = WW^{\rm T} + I \sigma^2_\varepsilon$, with $W\in \mathbb{R}^{p\times p^*}$ and $p^*<p$.
This allows us to use two latent vectors $z^t\sim \text{N}(0, I_{p^*})$ and $\xi^t\sim \text{N}(0, I_{p} \sigma^2_\varepsilon)$, and obtain $\varepsilon^t = W z^t + \xi^t \sim \text{N}(0, \Sigma_\varepsilon)$.

We introduce the following matrix notation $Y=[y^{ \mathcal T} \cdots  y^{d+1} ]^{\rm T} \in \mathbb{R}^{ ( \mathcal T-d)\times p}$, \\$\bar C=[C^{(1) }\cdots C^{(d) }]^{\rm T} \in \mathbb{R}^{ (pd)\times p}$, $X= \begin{bmatrix}
(y^{ \mathcal T-1})^{\rm T} & \ldots & (y^{\mathcal T-d})^{\rm T}\\
 \vdots & \ddots & \vdots \\
(y^{d})^{\rm T} & \cdots & (y^{1})^{\rm T}
\end{bmatrix} \in \mathbb{R}^{ ( \mathcal T-d)\times (pd)}$ and $Z = [z^{\mathcal T}\ldots z^{\mathcal  T-d} ] \in \mathbb{R}^{( \mathcal T-d) \times p^*}$ to write the likelihood in compact form:
\[\label{eq:likelihood}
 \mathcal{L}(Y,X,Z ; \bar C, W) \propto
(\sigma^2_\epsilon)^{-(\mathcal T-d)p/2} \exp( -  \frac{1}{2\sigma^{2}_\varepsilon}  \|Y- X  \bar C -  Z W^{\rm T} \|^2_F) \exp( - \frac{\|Z\|_F^2}{2}).
\]
\begin{remark}
The above likelihood function is also suitable for modeling multiple time series \\ $[Y^{(1)},X^{(1)}],\ldots, [Y^{(S)},X^{(S)}]$ based on a regression model with shared $\bar C$. In that case, one uses matrices $Y= [Y^{(1)} \cdots Y^{(S)}] \in \mathbb{R}^{(\sum_s \mathcal T_s -Sd)\times p} $, $X = [X^{(1)} \cdots X^{(S)}] \in \mathbb{R}^{ (\sum_s\mathcal T_s - Sd)\times (pd)} $, and adjusts the dimensions of other matrices accordingly.
\end{remark}

Next, we incorporate the prior information that $G$ should be sparse and nearly connected.
Consider the \textit{undirected} version of $G$, denoted by $\bar G$; that is, $\bar G=\{V,E_{\bar G}\}$, with $(i,j)\in E_{\bar G}$ if and only if at least one of $(i\to j)$ or $(j\to i)$ is in $G$]. We  assume that $\bar G$ can be covered by $m$ spanning trees,
\[
 \bar G \subseteq \bar T= \bigcup_{l=1}^m T^l,
\]
where the union and subset signs are shorthand for $E_{\bar G}\subseteq \bigcup E_{T^l}$ for notational convenience. Recall that a spanning tree $T^l$ is the smallest connected graph containing $p$ nodes with $(p-1)$ edges. Further, since we assume that it is ``connected'', for any two nodes $i$ and $j$, there is a set of edges in $E_{T^l}$ to form a path to them together.

\noindent\textbf{Graph-based Gaussian Prior Distribution with Further Edge Selection:}
To incorporate this structural assumption into the model, we use the following graph-based Gaussian prior distribution:
\begin{align}
\label{eq:direct_graph_reg}
& C_{i,j}^{(k)} \stackrel{indep}\sim \text{N} (0,  r_k \eta_{i,j}  \sigma^2_\varepsilon A_{\bar T:i,j}) , \\
\end{align}
where each $A_{\bar T} = (A_{\bar T:i,j})_{i,j=1,\cdots,p}$ is the adjacency matrix of the union of trees, with $A_{\bar T:i,j}=A_{\bar T:j,i}=1$ if $(i,j)\in \bar T$, and $A_{\bar T:i,j}=0$ otherwise; further, we fix $A_{\bar T:i,i}=1$.

When $A_{\bar T:i,j}=0$, the above distribution would be degenerate at point mass $C_{i,j}^{(k)}=0$. Further,  we use $\eta_{i,j}\ge 0$ and $r_k\ge 0$ to adjust for the varying scales of coefficients over $(i,j)$ and $k=1,\cdots, d$. Note that if $\eta_{i,j} = \eta_{j,i} = 0$ exactly and $(i,j)\in \bar T$, then $\bar G$ would be a disconnected subgraph of $\bar T$; if $\eta_{i,j}=0$ exactly, $(i,j)\in \bar T$, but $\eta_{j,i} \gg 0$, then $G$ would correspond to a directed graph. Therefore, the above graph-based Gaussian prior is quite flexible. On the other hand, to facilitate the computation of the posterior distribution, we will use strictly positive $\eta_{i,j}$ and $r_k$, and rely on a continuous shrinkage prior distribution to have some $\eta_{i,j}\approx 0$ and $r_k\approx 0$.
\(
& r_k \stackrel{iid}\sim \text{IG}(a_k,b_k), \\
& \eta_{i,j}\sim \text{IG}(\alpha_\eta,\beta_{i,j}), \; \beta_{i,j}\sim\text{Exp}(\gamma_\eta).
\)
where IG is the inverse-gamma distribution, and both inverse-gamma and exponential use scale parameterization.
The hierarchical prior on $\eta_{i,j}$ is equivalent to a generalized Pareto prior $\pi_0(\eta_{i,j})\propto (1+\eta_{i,j}/\gamma_\eta)^{-(1+ \alpha_\eta)}$. Since the true order of lags in the VAR model is unknown, we use a large value for the lag order $d$ and make the scale $b_k$ increasingly close to zero for larger values of $k$, as described at the end of this section.

\begin{remark}
We induce sparsity in \eqref{eq:direct_graph_reg} through the following two routes: we first select a connected and undirected graph $\bar T$ via binary $A_{\bar T}$, then we further select a subset of edges corresponding to $G \subseteq T$ via continuous shrinkage on $\eta_{i,j}$.
\end{remark}

For the parameters related to measurement error, we use
\(
 \sigma^2_\varepsilon \sim \text{IG}(\alpha_\sigma,\beta_\sigma),\qquad W_{i,j}\sim  \text{N}(0,\gamma_W).
\)
 We defer the specification of all the hyper-parameters to the end of this section.

\noindent\textbf{Prior Distribution for the Union of Trees:} In $\bar T$, each tree $T^l$ needs to satisfy the following constraints: (i) there are $(p-1)$ edges in $T^l$, (ii) $T^l$ needs to be connected.

Next, we assign a prior distribution for the union of trees $\bar T=\bigcup_{l=1}^m T^l$. We use the following discrete probability distribution that varies with the number of edges $|E_{\bar T}|$:
\[
\pi_0(\bar T) \propto {\lambda^{|E_{\bar T}|}},
\]
where the probability is normalized over all possible unions of $m$ spanning trees, and $\lambda>0$. It is not hard to see that if $\lambda>1$, we would encourage the $T^l$'s to have fewer overlapping edges; and if $\lambda<1$, we would favor more overlapping edges and consequently higher sparsity in $\bar C$.

A nice property of this prior distribution is that it allows two or more component trees to be identical $T^l=T^{l'}$, which is more likely to occur a priori when $\lambda<1$, compared to when $\lambda>1$. Since we do not know the number of trees to cover $\bar G$, we again set a large $m$, and rely on the above prior distribution with $\lambda<1$ to reduce the effective number of covering trees.

Another nice property is that the conditional prior probability for a component tree given the others is factorizable over the edges:
\(
\pi_0(T^l \mid \{T^{k}\}_{\text{all }k\neq l}) \propto {\prod_{(i,j)\in T^l} \big\{  \lambda {1[  (i,j)  \not\in \cup_{k\neq l}T^{k}]} +  1[ (i,j)  \in \cup_{k\neq l}T^{k}] \big\}}.
\)
This allows us to develop a tractable algorithm to update the component trees.

\noindent\textbf{Choice of the Hyper-parameters: }
Next, we specify the hyper-parameters mentioned above. First, we standardize each vector $(y^1_j, \ldots, y^\mathcal T_j)$, so that it has sample mean $0$ and sample variance $1$. This allows us to set the noise variance roughly on the same scale, $\sigma^2_\varepsilon \sim \text{Gamma}^{-1}(2,1)$ and $\gamma_W=1$. Next, for the generalized double Pareto distribution, we follow \cite{armagan2013generalized} and use $\alpha_\eta=3$ and $\gamma_\eta=0.001$ to balance between sparsity and tail-robustness. To regularize the order of autoregression, we use $a_k=3$ and $b_k=2\cdot 0.1^k$, corresponding to increasingly smaller prior mean $\mathbb{E}r_k = 0.1^k$ and variance $\mathbb{V}r_k = 0.1^{2k}$ as $k$ increases. For the union of trees prior distribution, we empirically find that having $\lambda$ adaptive to the length of the time series $\mathcal T$ is effective to control the number of edges $|E_{\bar T}|$, and we use $\lambda = 0.1^{\mathcal T}$ in this article. For the parameter dimensions, we use $d=m=10$.

\subsection{Arboricity, Tree Rank and Sub-graph Sparsity}

\textbf{Estimating the Granger causality graph:} Using the posterior sample, we can form an estimate of the graph $G$ via $A_G = A_{\bar T} \circ A_{\eta}$, with $A_{\eta:i,j} = 1(\eta_{i,j}\ge \delta)$ based on some threshold $\delta$.
To minimize the potential sensitivity in choosing  $\delta$, we select the one that has almost no impact on the model goodness-of-fit, measured by the Mean Squared Error (MSE). Let $\text{MSE}(\bar C) = \|Y-X \bar C\|^2_F/[(\mathcal T-d)p]$, and $\bar C_\delta$ be thresholded matrix, with $C^{(k)}_{\delta:i,j} = C^{(k)}_{i,j} 1(\eta_{i,j}\ge \delta)$, for each $\bar C$, we choose a maximal $\delta$ such that:
$|\text{MSE}(\bar C) - \text{MSE}(\bar C_\delta)|/\text{MSE}(\bar C)   \le \tilde\epsilon,$
with $\tilde\epsilon$ a small value (we use $\tilde \epsilon = 0.01$ in this article).

\begin{remark}
 A conceptually simpler solution could be obtained with a Bernoulli prior distribution on each element of matrix $\eta=\{ \eta_{i,j}\}_{\text{all} (i,j)}$, for which one could directly obtain an estimate of $G$ via $A_{G}=A_{\bar T} \circ \eta$. However, compared to a discrete model on $\eta$, the continuous shrinkage model gives rise to simpler computations ---  we will be able to integrate out $\bar C$ and rely on some fast tree sampling algorithm to update $\bar T$.
\end{remark}

Next, we discuss the consequences of covering $G$ with $m$ trees. First, note that the smallest number of trees covering $\bar G$ is less or equal to $m$. This is a summary statistic known as ``arboricity''.
\(
 \arg\min_{m'}  \{ (T^1,\ldots, T^{m'}) : \bar G \subseteq
\bigcup_{l=1}^{m'} T^l \}.
\)
We use the above for prior regularization, and call it the ``tree-rank''. It corresponds
to the number of independent factors (spanning trees) that form the basis of a graph \{for rigorous definitions of independence in graphs and bases, see \cite{murota1998discrete}\}. As the name implies, the tree-rank shares a similar range to a matrix-rank.
\setcounter{theorem}{0}
\begin{theorem}\label{thm:tree_rank}
For an undirected graph $\bar G$ with $p$ nodes, $1 \le \text{Tree-Rank}(\bar
G)\le p-1$.
\end{theorem}
Therefore, {analogously to imposing a low-rank constraint on matrices, a low tree-rank controls the complexity of the graph $\bar G$. On the other hand, a key difference from the matrix case is that a low tree-rank $m^*$ automatically ensures a certain level of sparsity, since $|E_{\bar G}|\le m^*(p-1)$.}
Further, the tree-rank also induces sparsity in {\em every} sub-graph of $\bar G$, as shown in the following classical result.
\begin{theorem}\cite{nash1964decomposition}
\[\label{eq:tree_rank}
\text{Tree-Rank}(\bar G)= \max_{H\subseteq \bar G} \left\lceil \frac{|E_H|}{|V_H|-1} \right\rceil.
\]
\end{theorem}
Therefore, with $\text{Tree-Rank}(\bar G)\le m$, we obtain that every subgraph $H\subset\bar G$ has at most $m(|V_H|-1)$ edges. That is, the tree-rank gives a much stronger control on the sparsity of $\bar G$.

\begin{remark}
The above theorem is very general and all undirected graphs (including small-world and scale-free graphs) satisfy this equality. In Appendix B, we provide an algorithm to estimate the tree-rank of a graph.
\end{remark}

\begin{figure}[H]
  \centering
  \subfloat[A sparse and connected graph having a low tree-rank.]{
			  \includegraphics[width=.3\textwidth]{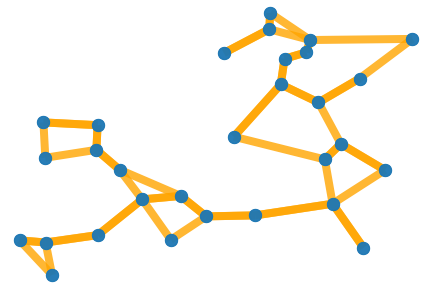}}\;\;
  \subfloat[The graph in (a) can be covered by two spanning trees (red and blue), each is a connected graph having only $(p-1)$ edges. ]{
  			\includegraphics[width=.3\textwidth]{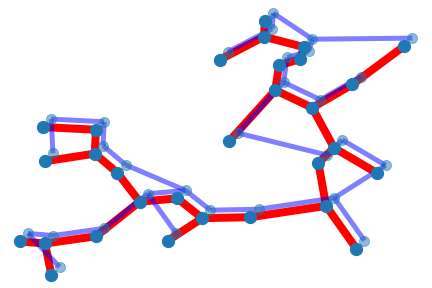}
	}\;\;
	  \subfloat[Another graph having a low matrix rank in the adjacency matrix. The graph is disconnected, and in this case, is dense in each component.]{  			\includegraphics[width=.3\textwidth]{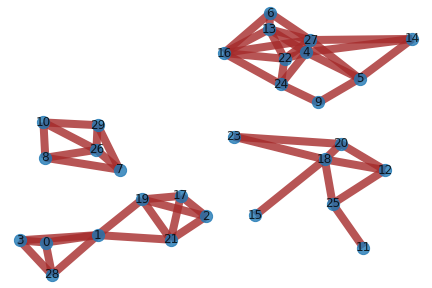}
	}
  \caption{Illustration of the low tree-rank graph modeling idea. \label{fig:low_tree_rank}}
\end{figure}

An illustration of the low tree-rank modeling idea is depicted in Figure \ref{fig:low_tree_rank}, which shows how a sparse and connected graph can be covered by two spanning trees, each being the smallest connected graph for $p$ nodes. In addition, it shows a fundamental difference between a graph of low {\em tree-rank} and a graph of low {\em matrix-rank} of its adjacency matrix: the former is connected and sparse, whereas the latter is disconnected and not guaranteed to be sparse.

\section{Gibbs Sampling for Posterior Computation}
\label{sec:computation}
Next, we derive an efficient and scalable algorithm for sampling the posterior distribution.

\subsection{Data Augmentation}

A challenge in updating the union of trees $\bar T$ is the quadratic term  $\bar C^{\rm T} X^{\rm T} X \bar C$ in the likelihood, which poses a combinatorial complexity when updating each edge of the tree. To address this issue,  we modify the Gaussian integral trick \citep{zhang2012continuous} and propose a new matrix Gaussian latent variable $U\sim \text{Matrix-N}[ (I \tilde m- X^{\rm T} X) \bar C, (I \tilde m- X^{\rm T} X), I\sigma^2_\epsilon ]$ with $U\in\mathbb{R}^{(dp)\times p}$, where we use $\tilde m = \|X^{\rm T} X\|+ \epsilon^*$ with $\|\cdot\|$ the spectral norm and $\epsilon^*=10^{-3}$ to ensure positive definiteness of the row covariance.
\(
&\Pi(U \mid X, \bar C, \sigma^2_\varepsilon) \\
& \propto (\sigma^2_\varepsilon)^{- (p^2d)/2} \exp \bigg\{ -\frac{1}{2\sigma^2_\varepsilon}  \text{tr} [U - (I \tilde m- X^{\rm T} X) \bar C]^{\rm T} (I \tilde m- X^{\rm T} X)^{-1}  [U - (I \tilde m- X^{\rm T} X) \bar C]  \bigg\}.
\)
Multiplying the above two yields the likelihood with augmented data:
\(
  \mathcal{L}(Y,X,U,Z; & \bar C, W, \sigma^2_\varepsilon)
 \propto (\sigma^2_\varepsilon)^{-(\mathcal T-d)p/2 - (p^2d)/2}  \exp( - \frac{\|Z\|_F^2}{2})
 \\
 & \cdot \exp \bigg[ -  \frac{1}{2\sigma^{2}_\varepsilon}
 \bigg(
   \|Y- Z W^{\rm T} \|_F^2 + \text{tr} [ U^{\rm T} (I \tilde m- X^{\rm T} X)^{-1} U]
 \\
 & \qquad  + \tilde m  \|\bar C\|_F^2
  -2 \text{tr} \big\{ \bar C^{\rm T} [X^{\rm T}(Y- Z W^{\rm T})+U] \big\}
 \bigg)
  \bigg].
\)
It is useful to note that the above likelihood is now conditionally independent for $C^{(k)}_{i,j}$ over all $(i,j)$. This allows us to develop an efficient collapsed Gibbs sampling algorithm.

\subsection{A Collapsed Gibbs Sampling Algorithm}
Based on the Gaussian prior distribution in \eqref{eq:direct_graph_reg} and $B=[X^{\rm T}(Y- Z W^{\rm T})+U]$, we can obtain the coefficient estimate via
\[
& (C_{i,j}^{(k)} \mid A_{\bar T:i,j}=1, .) \sim \text{N}  \big [  \frac{B_{i,j}^{(k)}}{ \tilde m+  (r_k \eta_{i,j})^{-1} },  \frac{\sigma^2_\varepsilon}{ \tilde m+  (r_k \eta_{i,j})^{-1} } \big ], \\
& (C_{i,j}^{(k)} \mid A_{\bar T:i,j}=0, .) =0,
\]
for all $(i,j,k)$ in a block.
Further, the Gaussian conjugacy allows us to integrate out those $C_{i,j}^{(k)}$ corresponding to $A_{\bar T: i,j}=1$ completely, leading to a marginal distribution of $A_{\bar T}$:
\(
&\Pi(A_{\bar T} \mid .) \propto
\prod_{(i,j)\in \bar T}  \lambda \tilde s_{i,j} \tilde s_{j,i},\\
&
\tilde s_{i,j}=\big[ \prod_{k=1}^d \frac{1}{(r_k \eta_{i,j} \tilde m+  1)^{1/2}}\big]
\exp \big[ \sum_{k=1}^d \frac{1}{2\sigma^2_\varepsilon}  \frac{[B_{i,j}^{(k)}]^2}{ \tilde m+  (r_k \eta_{i,j})^{-1} }\big].
\)
Therefore, conditioned on all the other trees $T^k: k\neq l$, we can update each tree via
\[
\Pi(T^l \mid \{T^k\}_{k\neq l}, .) \propto
\prod_{(i,j)\in T^l} \big\{  \lambda \tilde s_{i,j} \tilde s_{j,i} {1[  (i,j)  \not\in \cup_{k\neq l}T^{k}]} +  1[ (i,j)  \in \cup_{k\neq l}T^{k}] \big\},
\]
for $l=1,\ldots,m$.
Since the above is factorizable over the edges of $T^l$, we use the random-walk covering algorithm \citep{broder1989generating,aldous1990random,mosbah1999non} to sample from the above distribution. The algorithmic details can be found in the recent work of \cite{duan2022spectral}.

To update the parameters in the continuous shrinkage prior, we have
\[
 (r_k\mid .) &\sim \text{IG} ( \frac{|A_{\bar T}|_1}{2}+ a_k, \sum_{i,j} \frac{[C_{i,j}^{(k)}]^2}{2\eta_{i,j} \sigma^2_\varepsilon} + b_k), \\
 (\eta_{i,j}\mid .) &\sim \text{IG} ( \frac{d A_{\bar T:i,j}}{2}+ \alpha_\eta, \sum_{k} \frac{[C_{i,j}^{(k)}]^2}{2 r_k \sigma^2_\varepsilon} + \beta_{i,j}),\\
 (\beta_{i,j} \mid.) &\sim \text{Gamma} \big[ \alpha_\eta +1,  (\frac{1}{\gamma_\eta} +  \frac{1}{\eta_{i,j}})^{-1} \big],
\]
where all use the scale parameterization. To update the parameters related to the measurement error, we have
\(
(\sigma^2_\varepsilon \mid .) & \sim \text{IG} \bigg\{
\frac{(\mathcal T-d)p + |A_{\bar T}|_1 d}{2}  + \alpha_\sigma,
\frac{\|Y-X\bar C - ZW^{\rm T}\|_F^2}{2} +  \frac{\sum_{i,j,k}[C_{i,j}^{(k)}]^2}{2(r_k \eta_{i,j})}  + \beta_\sigma \bigg\},\\
(Z \mid .) & \sim \text{Matrix-N}[   (Y-X\bar C)W    (W^{\rm T}W/\sigma^2_\varepsilon + I)^{-1}/\sigma^2_\varepsilon, I, (W^{\rm T}W/\sigma^2_\varepsilon + I)^{-1} ],\\
(W \mid .) & \sim \text{Matrix-N}[  (Y-X\bar C)^{\rm T} Z    (Z^{\rm T}Z/\sigma^2_\varepsilon + I/\gamma_W)^{-1}/\sigma^2_\varepsilon, I, (Z^{\rm T}Z/\sigma^2_\varepsilon + I/\gamma_W)^{-1}].
\)
We provide empirical evidence in Appendix F that this algorithm enjoys rapid mixing of Markov chains.

\section{Consistency of Low Tree-Rank Vector Autoregression Models}
\label{sec:theory}

\subsection{Stability Condition}
\label{sec:stability}
The vector autoregressive process is stable if the evolving limit of the observations is finite as time $\mathcal T\to \infty$. Mathematically, the stability can be guaranteed  \citep{lutkepohl2005new,hamilton2020time} if
for any complex scalar $z\in \mathbb{C}:\ |z|\leq 1$,
\begin{align}\label{eq:general_stab}
      \det(I_p-C^{(1)} z-...-C^{(d)}z^d)\neq 0.
\end{align}
{Next, we derive an easy-to-verify sufficient condition.
Note that we can view $A(C,z):=C^{(1)}z+...+C^{(d)}z^d$ as a complex-valued and weighted adjacency matrix for a graph, where the weights correspond to the transition matrices $C^{(k)}, k=1,\cdots,d$. Since the graph Laplacian matrix is by construction positive semi-definite, we can enforce $ \det\{ I_p-A(C,z)\}>0$.}
\begin{theorem}\label{thm:stability}
Consider two transformed matrices of $C^{(1)},\cdots, C^{(d)}$ that are real-valued and symmetric:
\(
 (\tilde A^*)_{i,j}=  [ \sum_{k=1}^d  \{ \frac{C^{(k)}_{i,j}+C^{(k)}_{j,i}}{2} \} ^2 ] ^{1/2}, \;
 (\tilde A^{**})_{i,j} = [\sum_{k=1}^d  \{ \frac{ C^{(k)}_{i,j}+g_0 C^{(k)}_{j,i}}{2}\}^2 + (1- g_0^{2}) \{ \frac{C^{(k)}_{j,i}}{2}\}^{2} ] ^{1/2}
\)
for $i=1,\ldots,p$ and $j=1,\cdots,p$, with $g_0= -0.4 -0.61/d$. Then,
a sufficient condition for the vector autoregressive process \eqref{eq:var} to be stable is that for all $i$, the node strength
\(
\tilde D_i^{*}=\sum_{j=1}^p (\tilde A^{*})_{i,j} < 1/\sqrt{d}, \;\; \tilde D_i^{**}=\sum_{j=1}^p (\tilde A^{**})_{i,j}< 1/\sqrt{d}.
\)
\end{theorem}
\begin{remark}
This result holds for any vector autoregressive process, although it is particularly meaningful for low tree-rank and/or sparse models.  Since each node has few edges, hence most of $(C^{(k)}_{i,j},C^{(k)}_{j,i})$'s are zero, making the above condition easy to satisfy. A similar, but necessary condition  was derived in Proposition 2.2 (i) of \cite{Basu2015} that assumes \eqref{eq:general_stab} to be true.
Therefore, our new result shows that stability can be achieved via the sparsity condition.
\end{remark}

\begin{remark}
Note that for ease of computation, almost all estimation methods of VAR models do not impose the process stability constraint on the parameter estimates. Our algorithm follows this practice. On the other hand, in our collected posterior samples of $\bar C$, all of them satisfy the stability condition, even though the constraint was not enforced explicitly.
\end{remark}

\subsection{Consistent Estimation of the Transition Matrices}
\label{sec:consistency}
Next, we derive conditions for consistent estimation of the elements of the transition matrices, as the number of observations $ \mathcal T\to \infty$.
First, we rewrite the model in a linear regression form as
\begin{align}
    &Y=XC+\mathcal{E},
\end{align}where $\mathcal{E}:=[\varepsilon^\mathcal{T},\cdots,\varepsilon^{d+1}]^{\rm T}_{N\times p}$ is the error matrix.

We assume that observations are generated  with ground-truth $C_0= \{ C^{(1)}_0, \ldots, C^{(d)}_0\}^{\rm T}$, and associated ground-truth graph  $\bar{G_0}$. For ease of presentation, we denote with $N:= \mathcal T-d$ and use vectorized notation for $c=\text{vec}(C),y=\text{vec}(Y), c_0=\text{vec}(C_0)$, and $(i,j,k)$ as a shorthand for the corresponding vectorized single index $kp^{2}+jp+i$,

{Then, the likelihood function of model \eqref{eq:var} is given by}
        \(
                \mathcal{L}(y; c,\Sigma_\varepsilon) \propto \det(\Sigma_\varepsilon)^{-\frac{ \mathcal T-d}{2}}\exp [-\frac{1}{2} \{ y-(I_p\otimes X)c\}^{\rm T}(\Sigma_\varepsilon \otimes I_{T-d})^{-1} \{ y-(I_p\otimes X)c\} ].
        \)
The prior distribution on $c$ is
 $c \sim \text{N} ( 0,  \Phi )$,
with $\Phi =\text{diag} \{ r_k \eta_{i,j}  \sigma^2_\varepsilon A_{\bar T:i,j}\}$. This is based on the first line of \eqref{eq:direct_graph_reg}, where $c$ follows a Gaussian scale mixture prior.

The conditional posterior distribution is then given by
\begin{align}
 & (c \mid\ \Sigma_\varepsilon, r,\eta,s,A)\sim \text{N} \{ \hat{c}, (\hat{\Gamma}+\Phi^{-1}/N)^{-1}/N  \} \\
                &\hat{c}=(\hat{\Gamma}+\frac{\Phi^{-1}}{N})^{-1}\hat{\gamma},\\
                &\hat{\Gamma}=\Sigma_\varepsilon^{-1}\otimes X^{\rm T}X/N,\quad\hat{\gamma}=(\Sigma_\varepsilon^{-1}\otimes X^{\rm T})y/N.
        \end{align}

Next, we impose certain assumptions on $X, \mathcal{E}$ and some of the hyperparameters.

(A1) The value $r_k$, $s_l$ and $\eta_{i,j,k}$ are bounded from below by a constant that does not change with $N,p$, almost everywhere with respect to the posterior probability; whereas $A_{i,j}^l \in [\epsilon,1]$. As $\tau\to 0$, $\epsilon\to 0$ uniformly for any $N,p$.

(A2) $0<\inf_{N\geq 1}\lambda_{\min}(C_X)<\infty$, $0<\sup_{N\geq 1}\lambda_{\max}(C_X)<\infty$, where $C_X$ is the covariance matrix of each row of the data matrix $X$.

(A3) $||c_0||\leq K$, where $K$ is a positive constant.

(A4) $\lambda_{\min}(\Sigma_\varepsilon)>0,\lambda_{\max}(\Sigma_\varepsilon)<\infty$.

(A5) $p=o(N^{1/2})$.

(A6) The specified $d$, $p^*$ and $m$ are greater or equal to ground-truth values.

Assumption A1 ensures the boundedness of $||\Phi^{-1}||$, which plays an important role in the consistency proof. Assumptions A2 and A4 are standard ones for high-dimensional VAR models and ensure that $\lambda_{\min}(X^{\rm T}X/N)$ is bounded away from 0 and $\lambda_{\max}(X^{\rm T}X/N)$ is bounded above with high probability. In A1, we do not assume the true adjacency matrix to be known. On A5, we focus on the moderate dimension case $p^2/N\to 0$ for the theory.

Next, we establish that assuming $\bigcup_{l=1}^m T^l \nsupseteq \bar{G_0}$ holds,  the posterior probability of such $T_l$'s would go to 0 as $N\to\infty$. To do so, we compare posterior densities $\Pi(\Phi_\star\mid y,X)$ and $\Pi(\Phi_{\star\star}\mid y,X)$, where $\Phi_\star$ is the Gaussian scale parameter corresponding to a set of trees $\bigcup_{l=1}^m T_\star^l \supseteq \bar{G_0}$ and $\Phi_{\star\star}$ corresponding to a set of trees $\bigcup_{l=1}^m T_{\star\star}^l \not\supseteq \bar{G_0}$.

 \begin{theorem}\label{thm:posterior consistency}
Consider a stable VAR model with true parameter $c_0$ of tree-rank $m$ satisfying A1-A6; then, posterior consistency holds, i.e.,
\begin{align}
    &\Pi\{||c-c_0||>\eta \mid \Phi,y,X\}\to0, \text{ as }N\to\infty.
\end{align} Further, we have 
\begin{align}
    &\Pi(\Phi_{\star\star}\mid y,X)/\Pi(\Phi_{\star}\mid y,X)\overset{}{\rightarrow}0, \text{ as }N\to\infty,
\end{align}
where $\Phi_\star$ corresponds to a set of trees $\bigcup_{l=1}^m T_\star^l \supseteq \bar{G_0}$ and $\Phi_{\star\star}$ corresponds to a set of trees $\bigcup_{l=1}^m T_{\star\star}^l \not\supseteq \bar{G_0}$.
 \end{theorem}

 \begin{remark}
 The first result shows that the posterior distribution concentrates around the true parameter $c_0$, while the second one establishes a posterior ratio consistency for the trees covering the ground-truth graph $\bar G_0$ hence model selection consistency. In Appendix A, we further characterize the convergence rate when $\Phi_\star$ covers $\bar G_0$.
 \end{remark}

\section{Numerical Experiments on Synthetic Data}
\label{sec:simulation}

\subsection{Finite Sample Performance for Modeling Sparse Graphs}\label{sec:finite_perf}
We assess the finite sample performance of the model and the estimation procedure for a finite $\mathcal T$ varying between $400$ and $1200$, and for $p=30$ and $80$.  We experiment with two types of ground-truth Granger causal graphs $G_0$: (a) a low tree-rank one, and (b) a random sparse graph. The former is used to empirically show fast convergence of the posterior distribution, while the latter to assess the robustness of the posited model when the ground truth deviates from it.

For comparison, we also fit the generated data using: (i) a ``shrinkage only'' model, which is the Bayesian VAR model as described above except using continuous shrinkage only [by replacing $A_{T:i,j}$ with $1$ in \eqref{eq:direct_graph_reg}], (ii) a ``trees only'' model, a Bayesian model using a union of trees only [by replacing $\eta_{i,j}$ with $1$ in \eqref{eq:direct_graph_reg}], (iii) a VAR model with lasso regularization, and (iv) a VAR model with elastic net regularization. For (i), an alternative is to use a horseshoe prior regularization, although we find no clear difference in the results from the one based on the generalized Pareto prior distribution; hence, we only report the latter. For models (iii) and (iv), we use cross-validation to select the tuning parameters that control the amount of regularization.

For each $G_0$, with $d=3$, we randomly generate a transition matrix $C_0$ with $C_{0:i,j}^{(k)}$ from
$ \text{N}(0,1)$ if $(i,j)\in \bar G_0$, and  $C_{0:i,j}^{(k)}=0$ otherwise, then scale down to satisfy the stability condition in Theorem 3.
We use a covariance matrix $\Sigma_{\varepsilon}=\tilde\rho(0.5^{|i-j|})_{i,j = 1,\ldots,p}$, and then scale $\tilde\rho$ such that the signal-to-noise ratio $||C_0||_F/||\Sigma_\varepsilon||_F= 0.1$.

We assess the performance of the models on the following metrics:  (i) recovering $C_0$, by assessing the relative estimation error on the transition matrix  $||\hat{C}-C_0||_F/||C_0||_F$ with $\hat{C}$ the posterior mean (or, point estimate for lasso or elastic net); (ii) recovering the edges of $G_0$, by calculating the relative estimation error on the edges $\sum_{i,j}(A_{\hat G:i,j} \neq A_{G_0:i,j})/(p^2)$, with point estimate $\hat G$ corresponding to $\hat C_\delta$ as the thresholded version of $\hat C$, such that $|\text{MSE}(\hat C) - \text{MSE}(\hat C_\delta)|/\text{MSE}(\hat C)   \le \tilde\epsilon$. As described at the beginning of Section 2.2, for Bayesian models, we use the posterior mean of $\hat \eta$ during the thresholding procedure. It takes about $4$ minutes to run the MCMC algorithm for 1000 iterations at $p=30$, and $10$ minutes at $p=80$ on a quad-core laptop. Each setting is repeated 5 times and the average error rate and standard error are calculated.

\begin{figure}[H]
  \centering
  \subfloat[Relative estimation error for $C$ at $p=30$.]{\includegraphics[width=.48\textwidth]{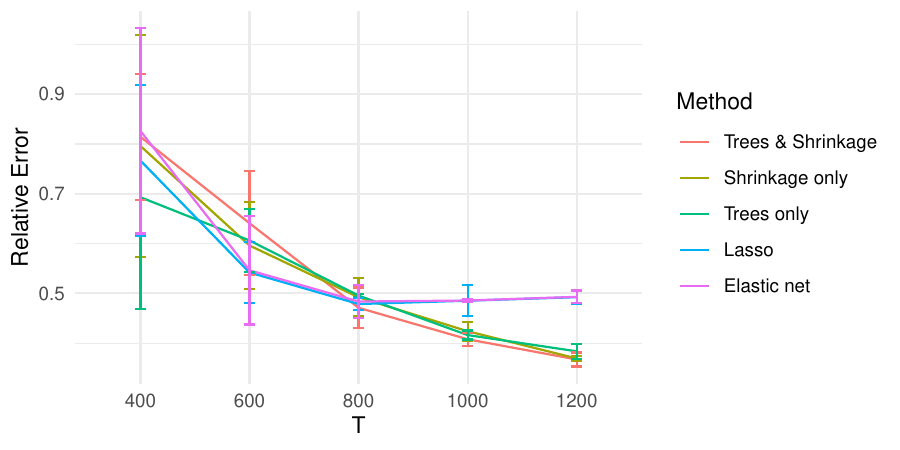}}\hfil
             \subfloat[Relative estimation error for $G$ at $p=30$.]{\includegraphics[width=.48\textwidth]{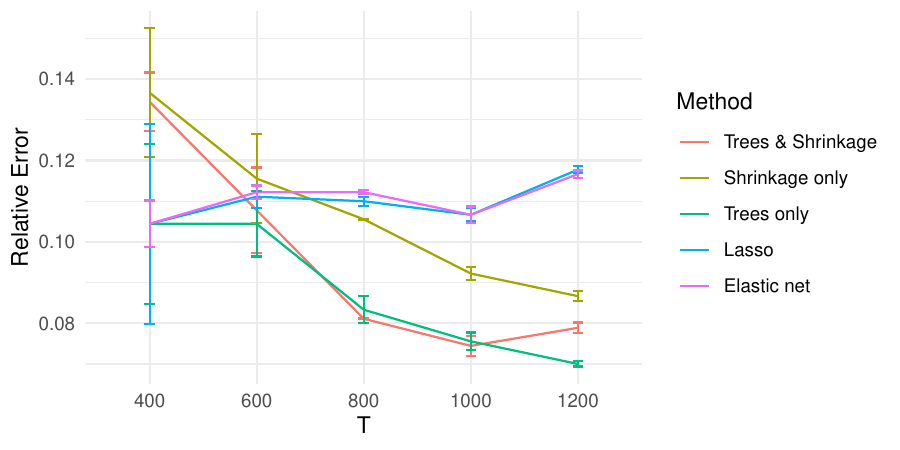}}\hfil
               \subfloat[Relative estimation error for $C$ at $p=80$.]{\includegraphics[width=.48\textwidth]{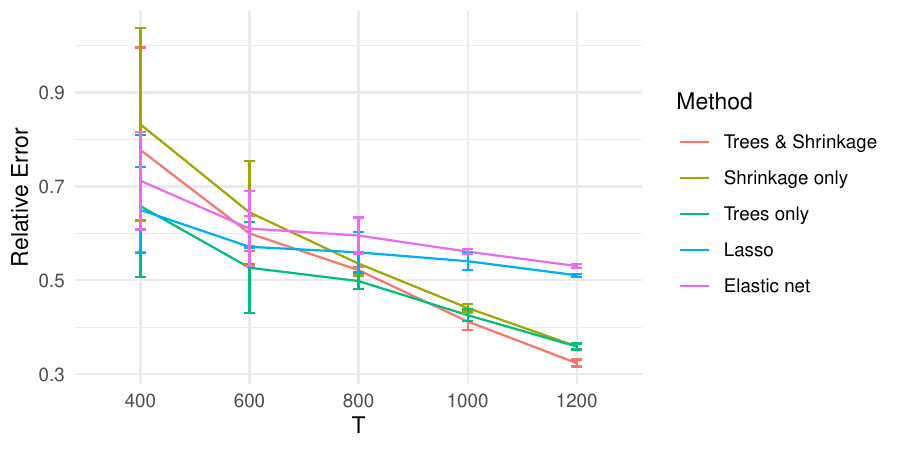}}\hfil
             \subfloat[Relative estimation error for $G$ at $p=80$.]{\includegraphics[width=.48\textwidth]{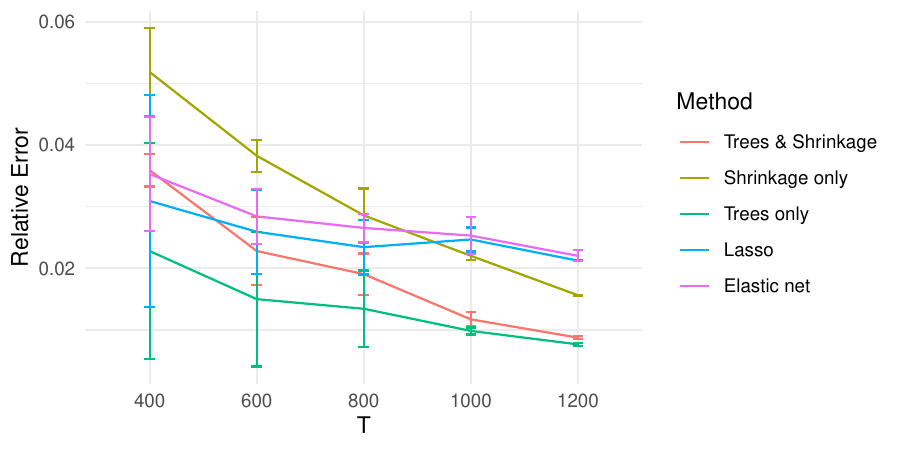}}\hfil
               \caption{Simulation results when the ground-truth graph $G_0$ has a low tree-rank at $2$. We compare five models: (i) the proposed model (Trees and Shrinkage), (ii) the Bayesian model  using the continuous shrinkage only (Shrinkage only), (iii) the Bayesian model using the union of trees only (Trees only), (iv) VAR regression with lasso regularization (Lasso), and (v) VAR regression with elastic net regularization (Elastic net).
                }
  \label{fig:senario1}
\end{figure}

We first consider the case where $G_0$ is indeed a graph of low tree-rank set to $2$ (Figure \ref{fig:senario1}). The trees only model shows the best performance, as it is one that corresponds to the true data-generating mechanism.  The proposed model has a very similar performance to the trees only model. The shrinkage only model shows slightly higher estimation errors. All three models show a rapid drop in the estimation errors as $\mathcal T$ increases. In comparison, the lasso and elastic net seem to have a relatively slow decrease of errors. 

\begin{figure}[H]
  \centering
  \subfloat[Relative estimation error for $C$ at $p=30$.]{\includegraphics[width=.48\textwidth]{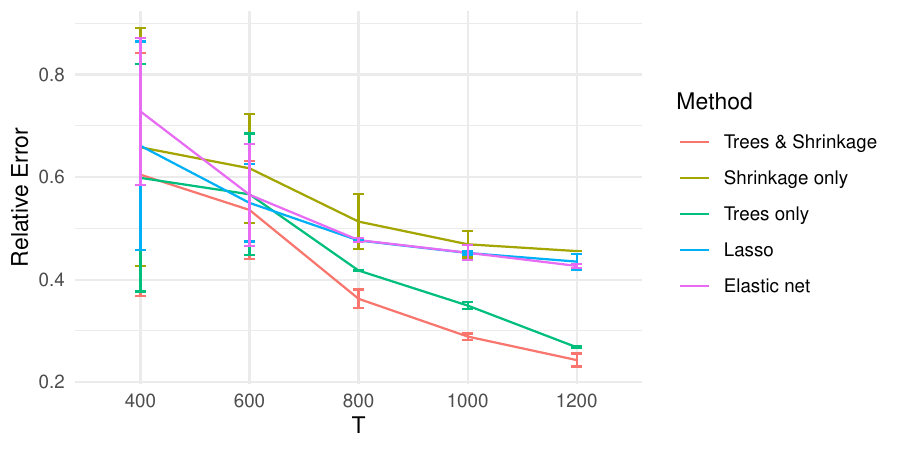}}\hfil
             \subfloat[Relative estimation error for $G$ at $p=30$.]{\includegraphics[width=.48\textwidth]{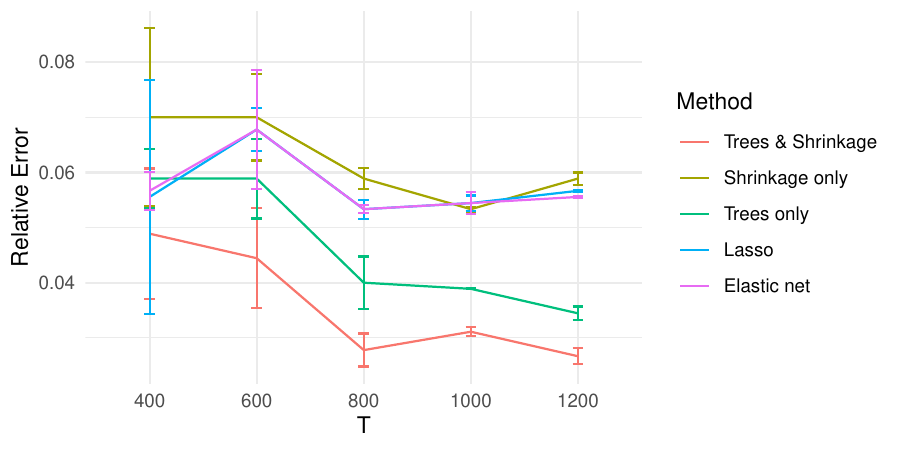}}\hfil
               \subfloat[Relative estimation error for $C$ at $p=80$.]{\includegraphics[width=.48\textwidth]{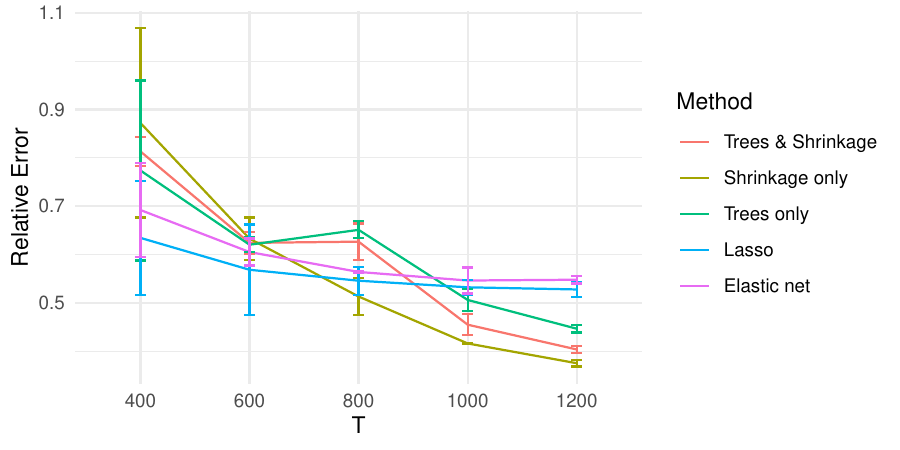}}\hfil
             \subfloat[Relative estimation error for $G$ at $p=80$.]{\includegraphics[width=.48\textwidth]{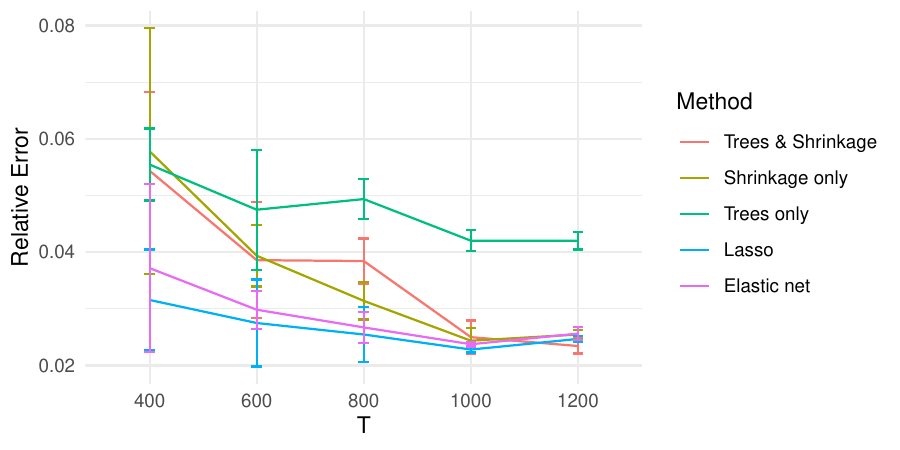}}\hfil
               \caption{Simulation results when the ground-truth graph $G_0$ has a 95\%
               sparsity.  }
  \label{fig:senario2}
\end{figure}

We then explore the case where $G_0$ is an unstructured sparse graph. We generate adjacency matrices $A_0$'s with about $5\% p (p-1)$ edges at random (Figure \ref{fig:senario2}). The results are very similar to the ones in the previous case, except that the trees only model now performs slightly worse than the proposed model. In Appendix G, we provide additional results for graphs at different edge densities.

In addition, since one could increase the threshold $\delta$ to have higher levels of sparsity in the graph estimate (although with greater compromise in the goodness-of-fit at a higher MSE), we evaluate the receiver operating characteristic curves for the above methods, and present them in Appendix E.

\vspace{-1cm}

\subsection{Modeling Relatively Dense Graphs}
 Our model was developed with the aim of modeling sparse graphs with a low tree-rank regularization; on the other hand, it can also be used for relatively dense graphs. Next, we illustrate that (i) the model can flexibly represent the underlying dense graph, provided the upper bound $m$ on the tree-rank is sufficiently large; (ii) even under an overly small $m$,  the constrained model still captures some important characteristics of the underlying graph.

\begin{figure}[H]
  \centering
             \subfloat[Ground-truth $p=80$.]{\includegraphics[width=.22\textwidth]{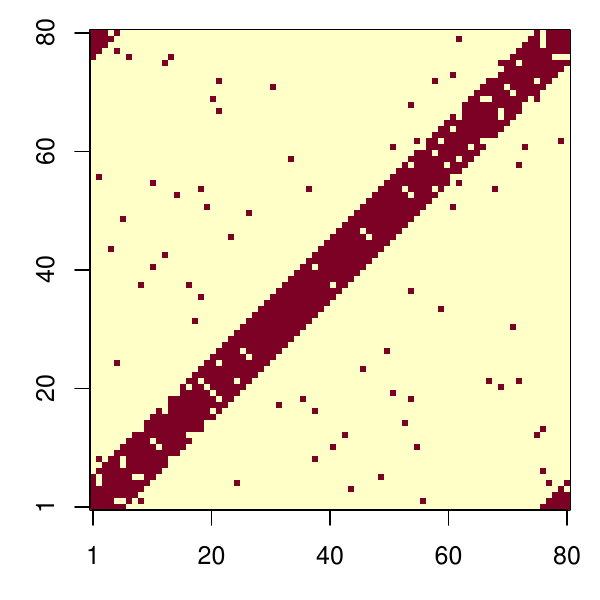}}\hfil
               \subfloat[Estimated graph using proposed model with tree-rank $\le$ 5.]
               {\includegraphics[width=.22\textwidth]{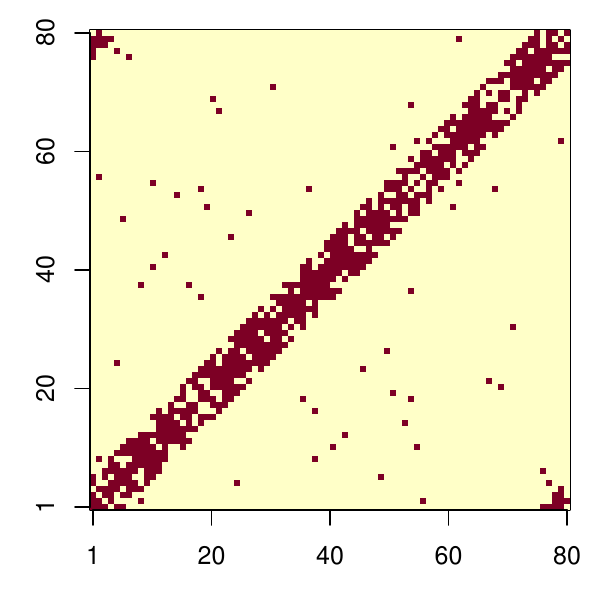}}\hfil
				\subfloat[Estimated graph using proposed model with tree-rank $\le$ 10.]{\includegraphics[width=.22\textwidth]{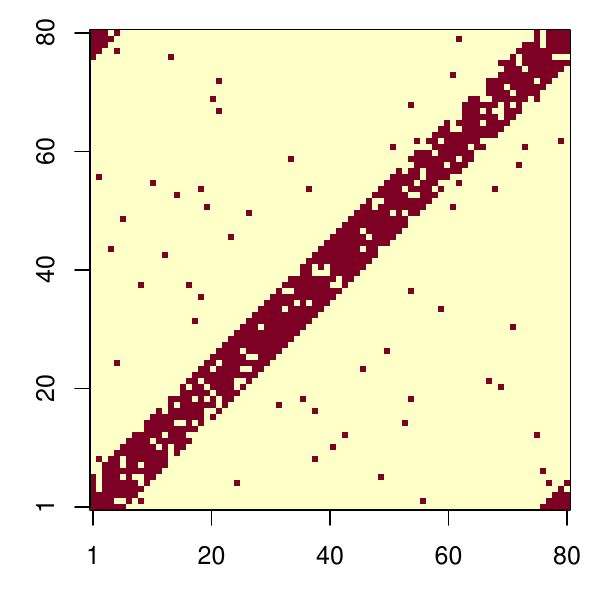}}\hfil
             \subfloat[Estimated graph using lasso.]{\includegraphics[width=.22\textwidth]{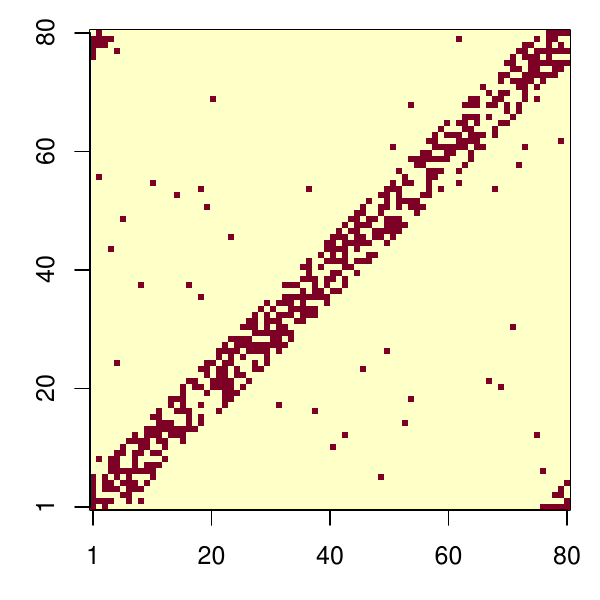}}\hfil
               \caption{Fitting the models when the ground-truth  $G_0$ is a small-world graph. \label{fig:small_world}}
\end{figure}

We first explore the case when $G_0$ is a small-world graph \citep{watts1998collective}. We generate $ G_0$ using the ``igraph'' function ``smallworld'' with a starting lattice of dimension $1$, neighborhood size $5$, rewiring probability $0.05$, and $p=80$. Based on $G_0$, we generate the transition matrix $C_0$ and the data as in Section \ref{sec:finite_perf}, and produce data over $\mathcal T=1200$ time points. Figure \ref{fig:small_world}(a) shows the ground-truth graph, and Panel(c) shows the estimated graph under tree-rank constraint $m\le 10$. Indeed, the estimated graph is very close to the ground truth. In addition, Panel (b) plots the estimated graph when the model is overly constrained with $m\le 5$. It can be seen that the estimated graph is clearly sparser than the ground truth, however, it still captures the ``small-worldness'', as those nodes indexed near 1 and those near 80 are directly connected by a few edges. Panel (d) shows a similar result when using lasso to fit the data.

We next consider the case when $G_0$ is a collection of several fully connected components (each component is a complete graph). We generate $G_0$ with component sizes of $20$, $30$ and $30$. Based on $G_0$, we generate the transition matrix $C_0$ and the data in the same way as in the last section, and produce data over $\mathcal T=1200$ time points. By Theorem 2, we can see that the grouth-truth tree-rank  of $G_0$ is $30$.

\begin{figure}[H]
  \centering
             \subfloat[Ground-truth $p=80$.]{\includegraphics[width=.22\textwidth]{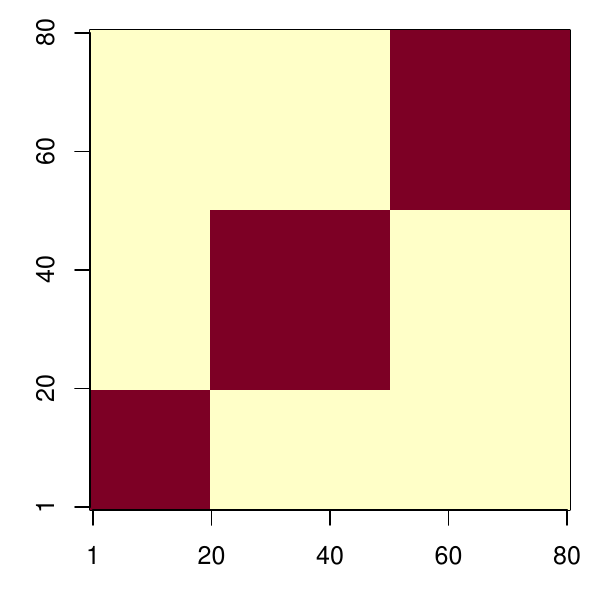}}\hfil
               \subfloat[Estimated graph using proposed model with tree-rank $\le$ 10.]
               {\includegraphics[width=.22\textwidth]{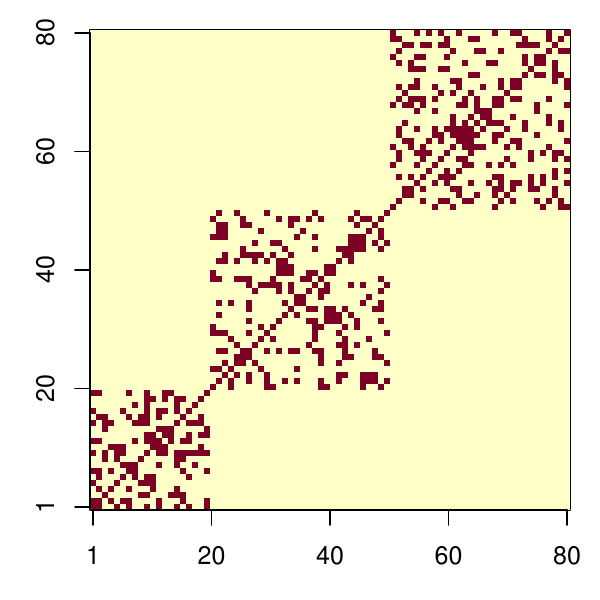}}\hfil
				\subfloat[Estimated graph using proposed model with tree-rank $\le$ 30.]{\includegraphics[width=.22\textwidth]{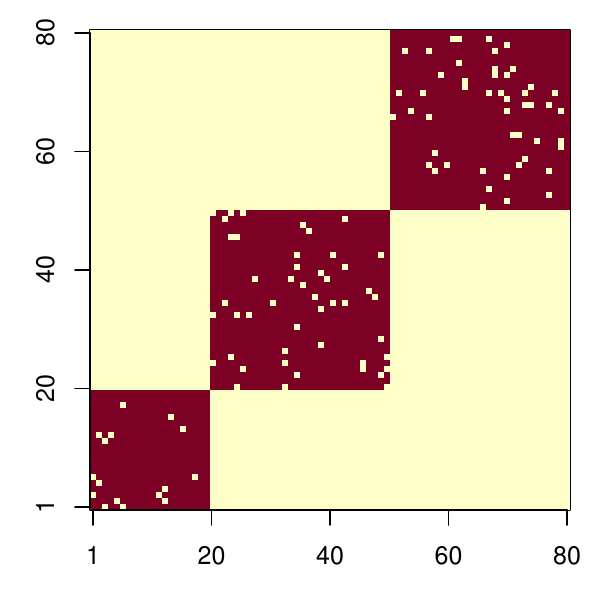}}\hfil
             \subfloat[Estimated graph using lasso.]{\includegraphics[width=.22\textwidth]{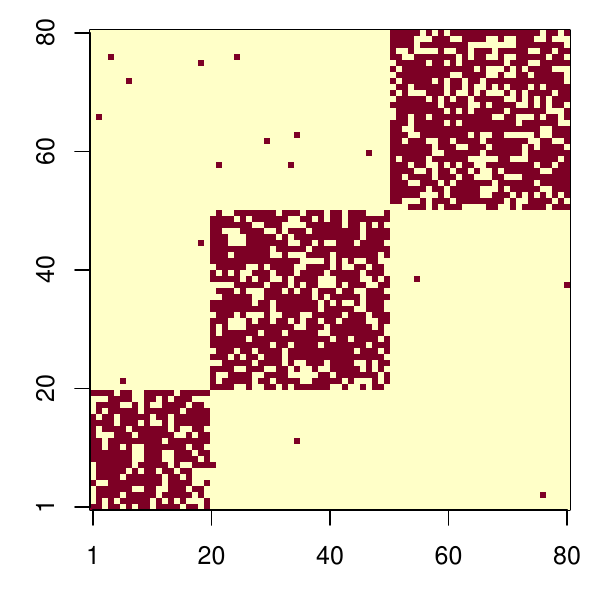}}\hfil
               \caption{Fitting the models when the ground-truth  $G_0$ is a graph with three fully connected components. \label{fig:cluster_graph}}
\end{figure}

Figure \ref{fig:cluster_graph}(a) shows the ground-truth graph, and Panel(c) shows the estimated graph under tree-rank constraint $m\le 30$. In addition, Panel (b) plots the estimated graph when the model is overly constrained with $m\le 10$. We can see again that the estimated graph is sparser than the ground truth, but captures the three-component structure. Panel (d) shows a similar result when using lasso to fit the data.

\section{Application to Brain Imaging Data}
\label{sec:application}

We employ the proposed model to analyze resting-state functional magnetic resonance imaging (fMRI) data from the Human Connectome Project. The fMRI data contain blood oxygen level-dependent (BOLD) signals for $\mathcal T=1200$. We use average BOLD signals in $68$ brain cortical regions of interest, according to the Desikan-Killarney atlas \citep{desikan2006automated}. We consider 468 subjects, each of whom has two scans taken at different times. We denote the first scan as the ``test" batch and the second scan as the ``retest" batch.

As our study focuses on reproducibility, we use the test batch as the training data for graph estimation and the retest batch as validation data to assess how many edge estimates can be reproduced. We fit our model by running the Markov chain Monte Carlo sampler for 2,000 iterations, discarding the first 1,000 as the burn-in period. We set the hyper-parameters according to the discussion in Section \ref{sec:model}, with $(m,d)=(10,10)$. For comparison purposes, we also fit sparse VAR models using (i) shrinkage only, (ii) trees only, (iii) lasso regularization, and (iv) elastic net regularization. It takes about 10 minutes to run the MCMC algorithm for each Bayesian model, and about 2 minutes to run the optimization algorithm for lasso or elastic net regularization on a quad-core laptop.

\begin{figure}[H]
  \centering
  \subfloat[Graph estimate using the proposed model. The graph has 268 edges.]
	{\begin{minipage}{0.3\textwidth}
		  \includegraphics[width=1\textwidth]{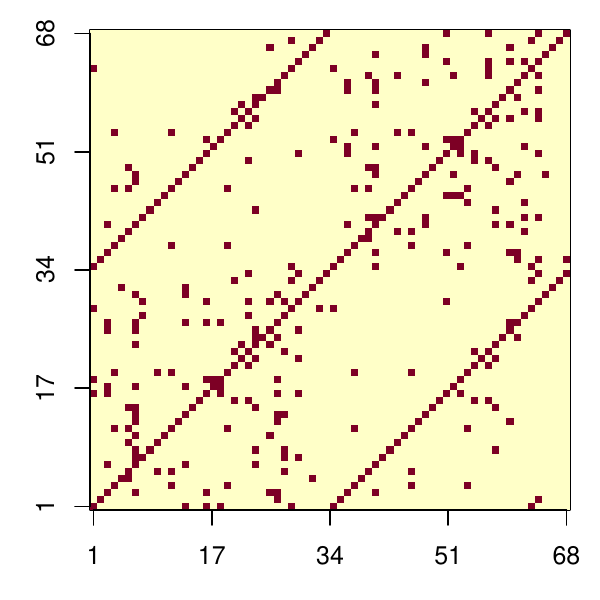}\\
  			\includegraphics[width=1\textwidth]{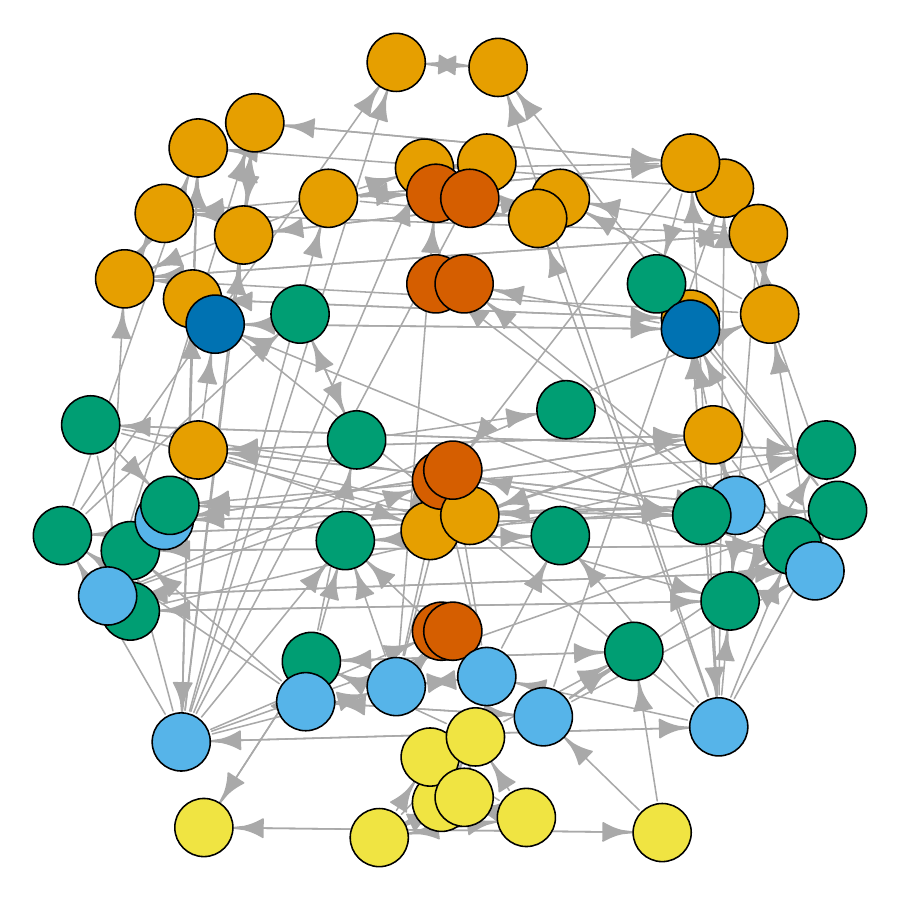}
	\end{minipage}
	}\;\;
  \subfloat[Graph estimate using the model with continuous shrinkage only. The graph has 405 edges.]{\begin{minipage}{0.3\textwidth}
		  \includegraphics[width=1\textwidth]{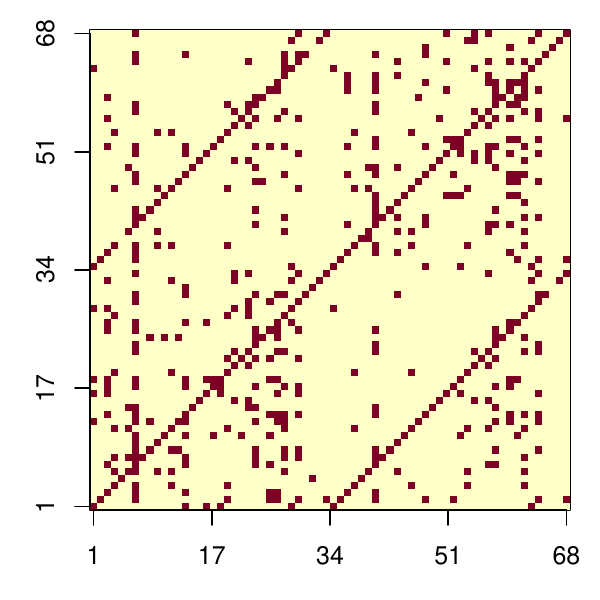}\\
  			\includegraphics[width=1\textwidth]{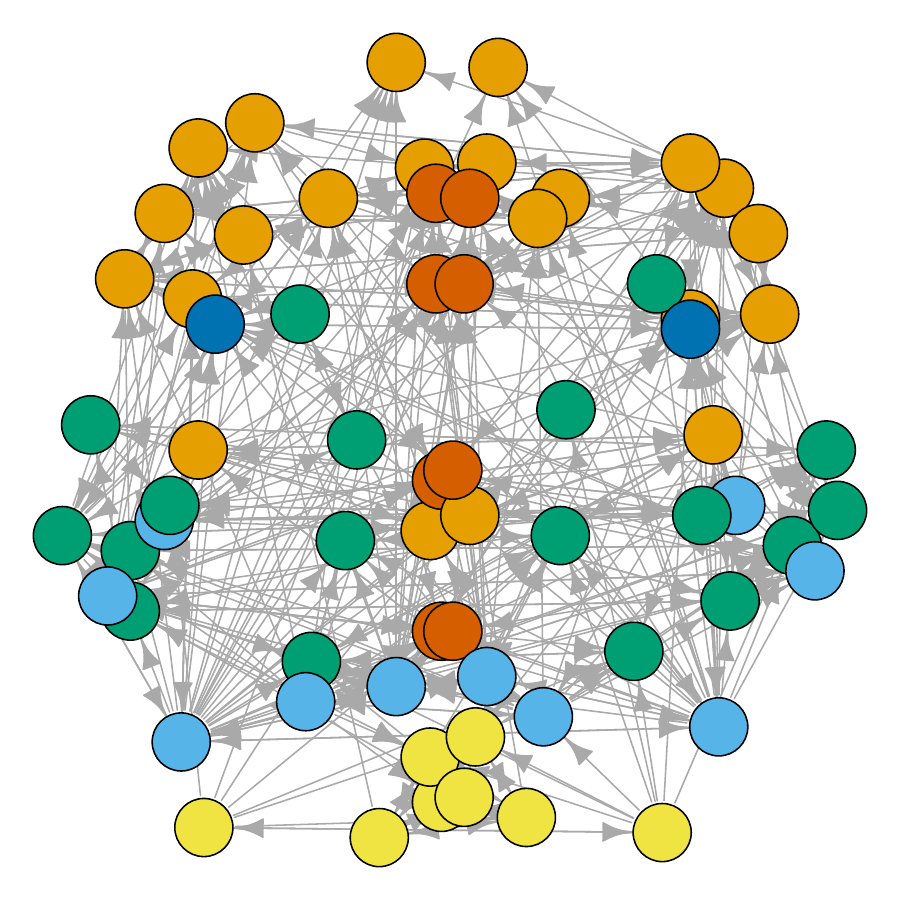}
	\end{minipage}
	}\;
 \subfloat[Graph estimate using the model with the union of trees only. The graph has 396 edges.]
	{\begin{minipage}{0.3\textwidth}
		  \includegraphics[width=1\textwidth]{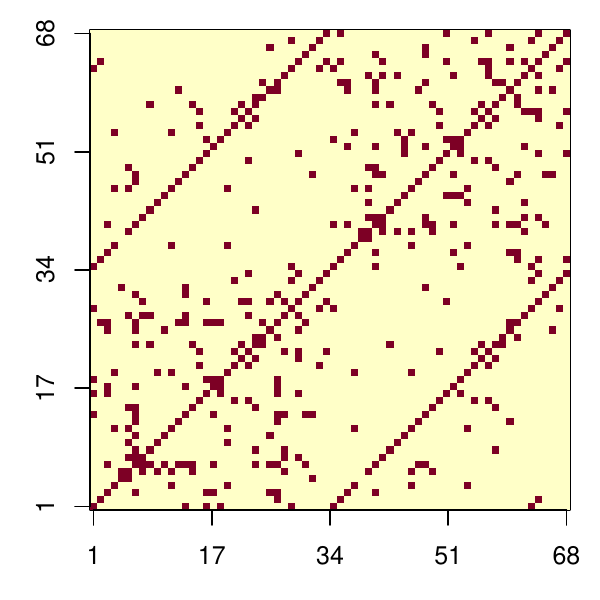}\\
  			\includegraphics[width=1\textwidth]{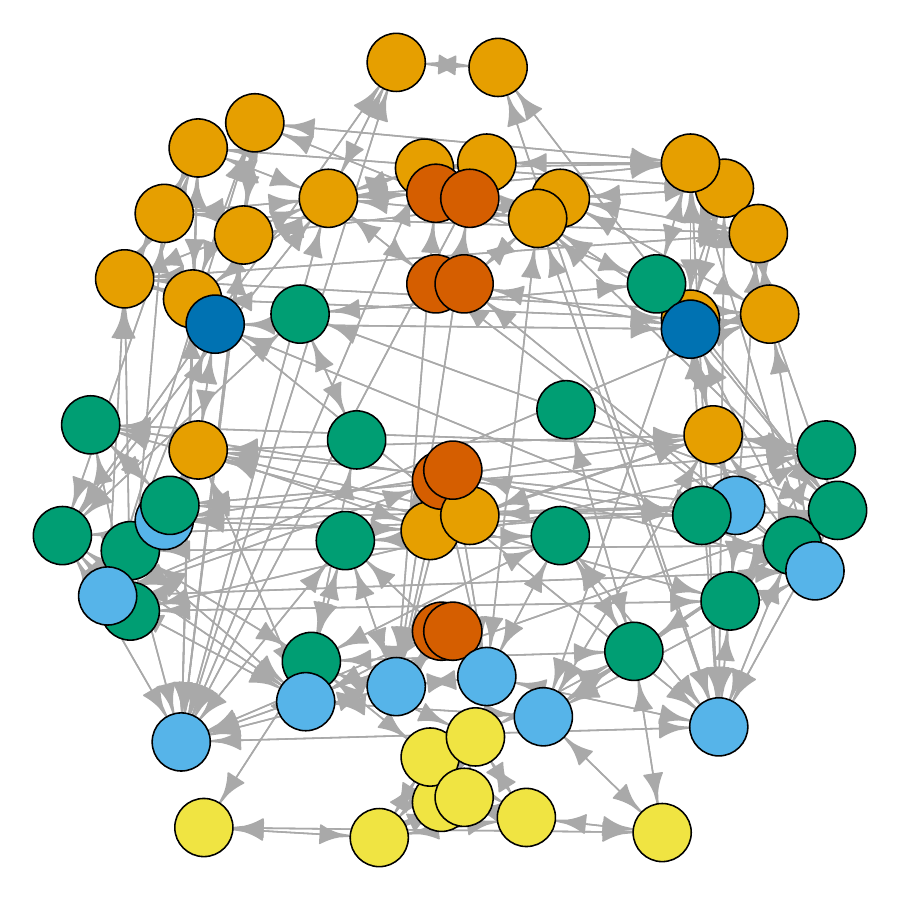}
	\end{minipage}
	}
  \caption{Graph estimates from the ``test'' batch of fMRI data. Nodes are plotted using the  Desikan-Killiany atlas node coordinates. }
  \label{fig:data_m}
\end{figure}

We form a point estimate $\hat G$ using the posterior mean (or the optimal value) $\hat C$, then threshold it using the procedure described in Section 2.2. For the Bayesian models, we use the posterior mean of $\eta$ in this step. As shown in Figure \ref{fig:data_m}, the proposed model shows the smallest number of edges in $\hat G$, followed by the trees-only model, and then the shrinkage-only model. The lasso and elastic net models have many more edges, which complicates interpretation (results shown in Appendix D).

\begin{figure}[H]
  \centering
  \subfloat[Graph estimate using the low tree-rank model with additional edge selection (proposed model). The graph has 261 edges.]
	{\begin{minipage}{0.3\textwidth}
		  \includegraphics[width=1\textwidth]{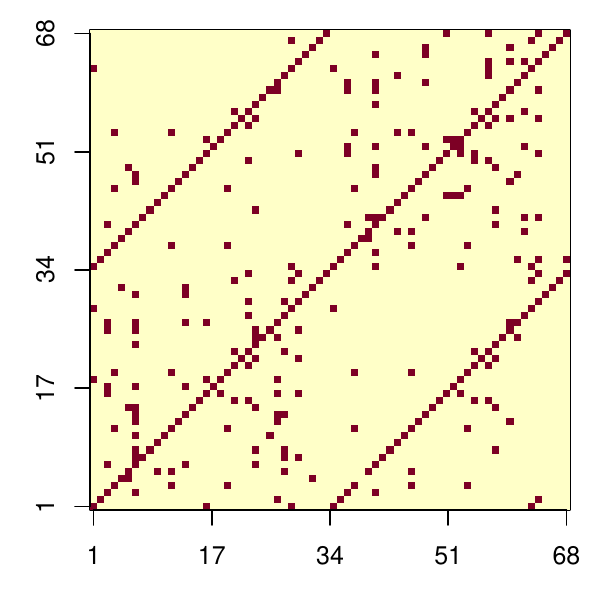}\\
	\end{minipage}
	}\;\;
  \subfloat[Graph estimate using the model with element-wise edge selection alone (via generalized Pareto shrinkage). The graph has 859 edges.]{\begin{minipage}{0.3\textwidth}
		  \includegraphics[width=1\textwidth]{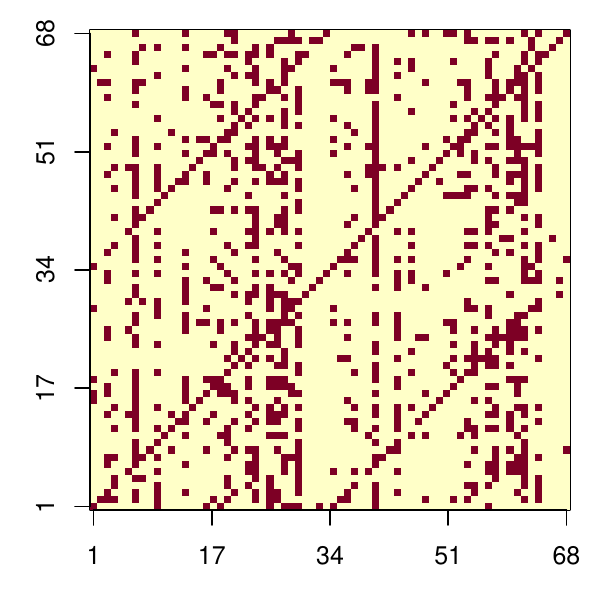}\\
	\end{minipage}
	}\;
 \subfloat[Graph estimate using the model with the union of trees prior alone. The graph has 398 edges.]
	{\begin{minipage}{0.3\textwidth}
		  \includegraphics[width=1\textwidth]{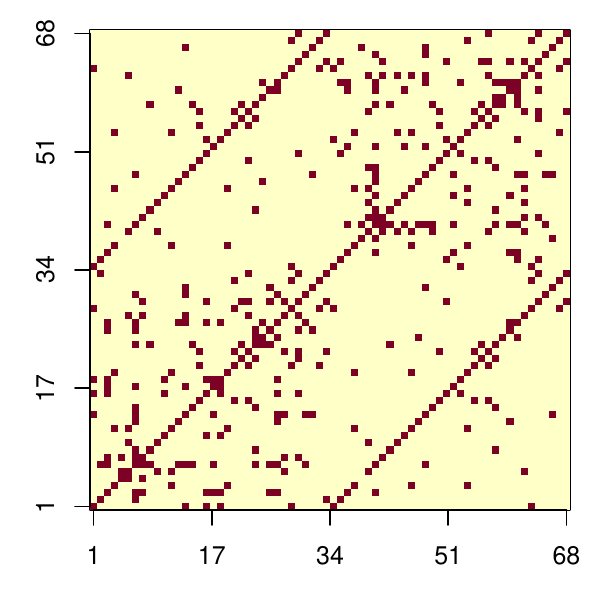}\\
	\end{minipage}
	}
  \caption{Graph estimates from the ``retest'' batch of fMRI data.}
  \label{fig:data_m2}
\end{figure}

We then fit these models to the ``retest" batch, using the same hyper-parameters, and show the results in Figure \ref{fig:data_m2}. Compared with Figure \ref{fig:data_m}, we can see that there is not much change in $\bar G$ between the ``test" and ``retest" for the proposed model and for the trees-only model; whereas the graph estimate from the ``retest" is much denser than the one from the ``test" for the shrinkage-only model.

To quantify the changes, we calculate the Jaccard index as a reproducibility score that compares the graph estimates in two batches for each of the five methods (Table \ref{Tab:error}). The proposed model and the trees-only model show the highest Jaccard index score, whereas the shrinkage-only model, lasso, and elastic net show much lower scores.

In addition, we explore calibrating the shrinkage-only model by increasing the prior penalty, so that it can produce a similar level of sparsity to that of our proposed model. To do this, we increase $\alpha_\eta$ to $30$ and reduce $\gamma_\eta$ to $0.0001$, obtaining $257$ edges in the graph estimate from the ``test" batch. Nevertheless, this calibrated model produces $648$ edges from the ``retest" data, and the Jaccard index is worse than the uncalibrated version of the shrinkage-only model.

\vspace{-1cm}

\begin{table}[!ht]
\centering
                \begin{tabular}{|l|l|l|l|l| l|}
                        \hline
                     Proposed   & Shrinkage only & Trees only  & Lasso  & Elastic net & Shrinkage only (calibrated) \\ \hline 
0.951 & 0.750 & 0.941 & 0.737 & 0.721 & 0.632
\\ \hline 
                \end{tabular}
                \caption{The Jaccard index comparing the two graph estimates from the ``test'' and ``retest'' batches of data. The proposed model and the trees only model show the highest score of Jaccard index. \label{Tab:error}}
               
        \end{table}

\begin{figure}[H]
\centering
  \subfloat[The probability estimate of $\Pi((i\to j)\in G\mid y )$ based on the ``test'' data.]
	{\begin{minipage}{0.4\textwidth}
		  \includegraphics[width=1\textwidth]{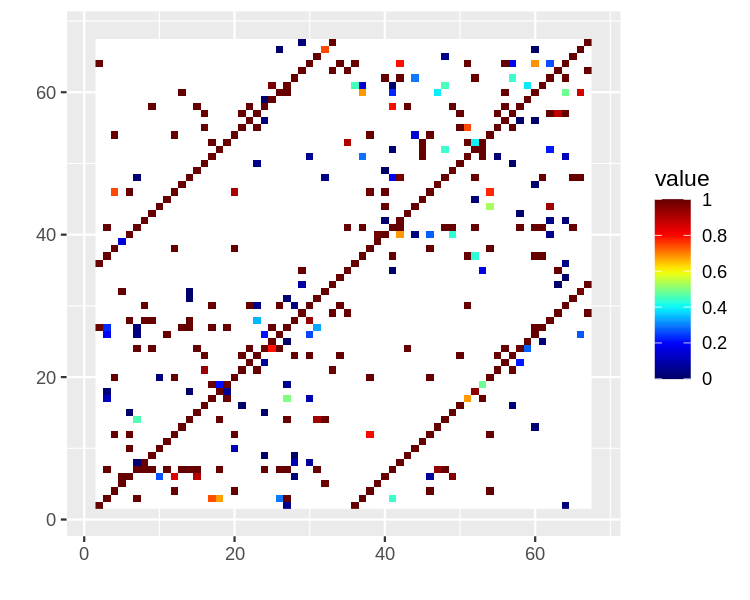}\\
	\end{minipage}
	}\;\;
  \subfloat[The probability estimate of $\Pi((i\to j)\in G\mid y )$  based on the ``retest'' data.]
  {\begin{minipage}{0.4\textwidth}
		  \includegraphics[width=1\textwidth]{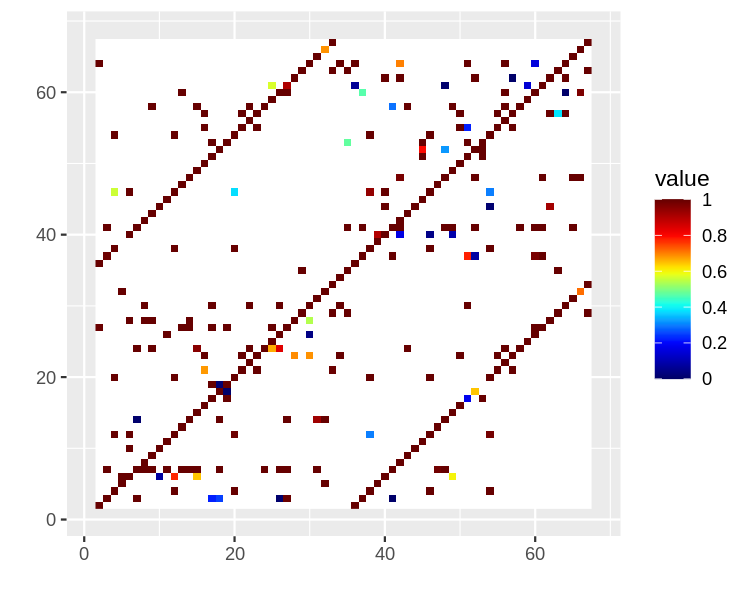}\\
	\end{minipage}
	}\\
	  \subfloat[The posterior distribution of the effective tree-rank in $G$ based on the ``test'' data.]
	{\begin{minipage}{0.4\textwidth}
		  \includegraphics[width=1\textwidth]{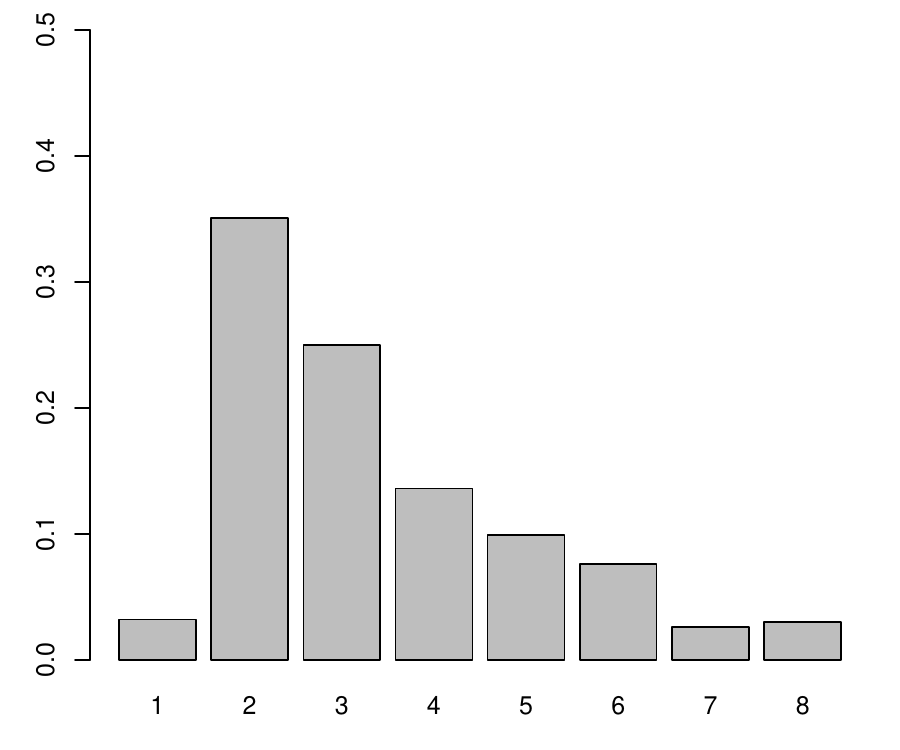}\\
	\end{minipage}
	}\;\;
	  \subfloat[The posterior distribution of the effective tree-rank in $G$ based on the ``retest'' data.]
  {\begin{minipage}{0.4\textwidth}
		  \includegraphics[width=1\textwidth]{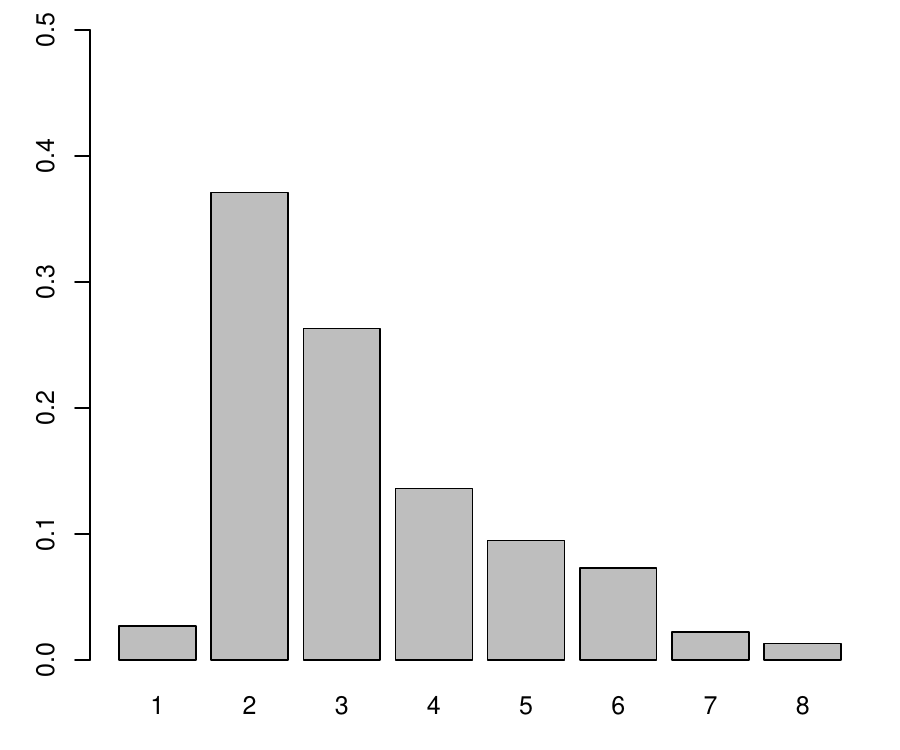}\\
	\end{minipage}
	}
  \caption{Uncertainty quantification on the graph estimates from the proposed model. \label{fig:uq}}
\end{figure}

Lastly, we use the posterior sample of $G$ and quantify the uncertainty associated with the point estimate $\hat G$. We estimate the probability $\Pi((i\to j)\in G\mid y )$, by calculating the proportion when $(i\to j)$ is included in the posterior sample point of $G$. As shown in Figure \ref{fig:uq}, most of the edges have relatively low uncertainty. Further, comparing the two batches, most of these edges of low uncertainty seem to appear in both of the graphs.

 \section{Discussion}
 \label{sec:discussion}
We introduce a tree-rank prior distribution to induce both near-full connectivity and high sparsity in the Granger causal network of a VAR model. We propose a fast algorithm for calculating the posterior distribution of the model parameters and establish posterior consistency of these estimates.

There are several interesting extensions to pursue in future work. The focus of this paper was a single connected, but highly sparse Granger causal network. However, there might be networks with relatively low tree-rank, which contain several small, dense sub-networks. To accommodate this structure, we could adopt the approach from \cite{Basu2019} and consider a more flexible ``low tree-rank plus sparse" structure. In addition, it will be of great interest to link the estimated Granger causal networks from fMRI to behavior traits and disease status. Considering that the proposed method has estimated surprisingly reliable networks from a relatively less reproducible imaging modality-fMRI \citep{zuo2019reliability}, we expect to get reliable and reproducible results in such analyses.  On the theoretical front, we have established the consistency of parameter estimates. It would be interesting to quantify the convergence rate, although characterizing the tree-covering probability poses a significant technical challenge. 

\bibliographystyle{plainnat}
\bibliography{ref}

\appendix

\section*{Appendix}

\section{Proofs}

\subsection{Proof for Theorem 1}

The lower bound is trivial. For the upper bound, note that $\bar G\subseteq
F$, with $F$ a complete graph $F=\{V, E_F\}$, with $E_F$ containing all possible
pairs of undirected $(i,j)$. Then consider the tree $\tilde T^l$ with edge
sets $E_{\tilde T^l}=\{ ( [l+1]_{p},1), ([l+2]_p ,2),\ldots, ([l+p-1]_p ,p-1)
\} $, where $[l+i]_p=l+i$ if $l+i\le p$, otherwise $[l+i]_p=l+i-p$ --- that
is,  $E_{\tilde T^l}$ corresponds to the $l$-diagonal elements (the diagonal
with $l$ row offset from the main diagonal ) in a $p\times p$ matrix.

Therefore, we can see that $\cup_{l=1}^{p-1} \tilde T^l$ includes all
the edges in $F$, therefore, the tree rank of $F$ is at most $p-1$, hence
so is for $\bar G$.

\subsection{Proof for Theorem 2}

For a complex matrix $A \in \mathbb{C}^{p \times p}$, $A^{*}$ denotes the conjugate transpose of $A$. We use $\overline{z}$ to denote the complex conjugate of a complex number $z$. The modulus of $z$ is denoted by $|z|$. For a complex matrix $A \in \mathbb{C}^{p \times p}$, $A_H$ denotes the Hermitian part of $A$: $A_H=2^{-1}(A+A^{*})$. Let $A(C,z):=C^{(1)}z+...+C^{(d)}z^d$. We use $B(z)=2^{-1}[A(C,z)+ A^*(C,z)]$ to denote its Hermitian part. Since  $B(z)$ is Hermitian, we define the Laplacian as $L_B(z):=D_B(z)-B(z)$ with $D_{B}(z)= \text{diag}\{\sum_{j=1}^{p}|B(z)_{ij}|\}_i$. Clearly $L_B(z)$ is Hermitian hence all eigenvalues of $L_B(z)$ are real, denoted by $\lambda_1 \leq \lambda_2\leq...\leq \lambda_p$.

For ease of notation, we omit $._B(z)$ for now. For any $w=(w_1,...,w_p)^T \in \mathbb{C}^p$, we have
\begin{align}
        &w^{*}Lw=w^{*}(D-B)w\\
        &= \sum_{i=1}^{p}\overline{w}_i (\sum_{j=1}^p|B_{ij}|) w_i - \sum_{i<j}\overline w_i B_{i,j} w_j
        - \sum_{j<i}\overline w_i B_{ij} w_j -  \sum_{i=1}^n \overline{w}_i B_{ii} w_i\\
        & \stackrel{(a)}= \sum_{i=1}^{p} |{w}_i|^2 (\sum_{j\neq i}|B_{ij}|)  -2 \sum_{i<j}\overline w_i B_{ij} w_j\\
                & =\frac{1}{2} \sum_{i=1}^{p}  \sum_{j\neq i} |B_{ij} |(\overline w_i - \frac{\overline B_{ij}}{|B_{ij}|}\overline w_j) ({w}_i -\frac{B_{ij}}{|B_{ij}|} w_j)\\
& =\frac{1}{2} \sum_{i=1}^{p}  \sum_{j\neq i} |B_{ij} | |{w}_i -\frac{B_{ij}}{|B_{ij}|} w_j|^2 \\
& \ge 0.
\end{align}
where $(a)$ is due to $B_{i,j}= \overline B_{j,i}$ and $|B_{i,i}|=B_{i,i}$.

Therefore, $L$ is positive semi-definite; and it is not hard to see that $L+\varepsilon I_p$ is strictly positive definite, for any $\varepsilon>0$.

For the complex matrix $W=I_p - A(C,z)$ to be positive definite, the sufficient and necessary condition is that its Hermitian part is  positive definite. That is:
\begin{align}
        &W_H=I_p-B\\
        &=(L+\varepsilon I_p)+(1-\varepsilon)I_p-D_B
\end{align}
should be positive definite. Since $D_B$ is diagonal, therefore, a sufficient condition is to have:
\begin{align}
        1-\varepsilon - (D_B)_{ii} \ge 0
\end{align}
for all $i=1,\ldots,p$. This is equivalent to
\begin{align}
\sum_{j=1}^{p}|\frac{1}{2}[A(C,z)+A(C,z)^{*}]_{ij}| \le 1-\varepsilon,
\end{align}
or
\begin{align}
\sum_{j=1}^{p}|C^{(1)}_{i,j} z+...+C^{(d)}_{i,j}z^d+ C^{(1)}_{j,i}\overline z+...+C_{j,i}^{(d)}\overline z^d| \le 2-2\varepsilon,
\end{align}

Taking $\varepsilon\to 0_+$, we have a sufficient condition for stability:
        \begin{align}
\sum_{j=1}^{p}|C^{(1)}_{i,j} z+...+C^{(d)}_{i,j}z^d+ C^{(1)}_{j,i}\overline z+...+C_{j,i}^{(d)}\overline z^d| < 2 \qquad \forall z\in \mathbb{C}:\ |z|\leq 1
\end{align}
for all $i=1,\ldots,p$.

Using the polar coordinate for $z= r \exp(\mathbb{I}x) $, where $r\in[0,1]$ and $\mathbb{I}$ is the imaginary unit. The above becomes:
        \begin{align}
\sum_{j=1}^{p}|\sum_{k=1}^d [C^{(k)}_{i,j} r^{k}\exp(\mathbb{I}kx)+ C^{(k)}_{j,i} r^{k}\exp(-\mathbb{I}kx)]| < 2 \qquad \forall z\in \mathbb{C}:\ |z|\leq 1
\end{align}

For each term on the left-hand side, it has
\begin{align}
& |\sum_{k=1}^d [ (C^{(k)}_{i,j}+C^{(k)}_{j,i}) \cos(kx)r^{k} + \mathbb{I}(C^{(k)}_{i,j}-C^{(k)}_{j,i}) \sin(kx)r^{k}|\\
&= \left\{[\sum_{k=1}^d  (C^{(k)}_{i,j}+C^{(k)}_{j,i})^{} \cos(kx)r^{k} ]^{2}+ [\sum_{k=1}^d(C^{(k)}_{i,j}-C^{(k)}_{j,i}) \sin(kx)r^{k}]^{2}
\right\}^{1/2} \\
& \stackrel{(a)}\le  \left\{\sum_{k=1}^d  (C^{(k)}_{i,j}+C^{(k)}_{j,i})^{2} r^{2k}\sum_{k=1}^d \cos^{2}(kx) + \sum_{k=1}^d(C^{(k)}_{i,j}-C^{(k)}_{j,i})^{2} r^{2k}\sum_{k=1}^d \sin^{2}(kx) \right\}^{1/2} \\
& = \bigg\{ [ \sum_{k=1}^d\cos^{2}(kx)+\sum_{k=1}^d\sin^{2}(kx) ]\sum_{k=1}^d [(C^{(k)}_{i,j})^2+(C^{(k)}_{j,i})^{2}] r^{2k}  \\
& \qquad+
 [ \sum_{k=1}^d\cos^{2}(kx)-\sum_{k=1}^d\sin^{2}(kx) ]\sum_{k=1}^d[2(C^{(k)}_{i,j})(C^{(k)}_{j,i})] r^{2k}
\bigg\}^{1/2} \\
& = \left\{d\sum_{k=1}^d [(C^{(k)}_{i,j})^2+(C^{(k)}_{j,i})^{2}] r^{2k}+ 2 [\sum_{k=1}^d\cos(2kx) ]\sum_{k=1}^d C^{(k)}_{i,j} C^{(k)}_{j,i}] r^{2k}
\right\}^{1/2}
\end{align}
where $(a)$ uses the Cauchy-Schwarz inequality, we denote:
\begin{align}
   g_{x}&= \frac{1}{d} \sum_{k=1}^d\cos(2kx) \\
   & = \frac{1}{2d} [1+ 2\sum_{k=1}^d\cos(2kx)]- \frac{1}{2d}\\
& = \frac{2\pi}{2d} D_d(2x) - \frac{1}{2d},
\end{align}
where $D_{d}(x)$ denotes the Dirichlet kernel, which has the maximum of $(2d+1)/(2\pi)$, and the minimum around $-c_0 (2d+1)/(2\pi)$ for $d\ge 10$, with $c_0\approx 0.2172$.  Taking $c_1=0.22$, we have $D_{d}(x)>-0.22(2d+1)/(2\pi).$ Slightly adjusting the constant, we have:
\begin{align}
-0.4- \frac{0.61}{d}< g(x)\le 1
\end{align}
for all $d\in\mathbb{Z}_+$, we denote  $g_0 = -0.4- {0.61}/{d}$.

Continuing on the inequality,
   \begin{align}
&  \left\{d\sum_{k=1}^d [(C^{(k)}_{i,j})^2+(C^{(k)}_{j,i})^{2}] r^{2k}+ d g_x [ \sum_{k=1}^d 2C^{(k)}_{i,j} C^{(k)}_{j,i}] r^{2k} \right\}^{1/2} \\
& = \left\{d\sum_{k=1}^d \left( [C^{(k)}_{i,j}+g_xC^{(k)}_{j,i}]^2 + (1- g^2_x) (C^{(k)}_{j,i})^{2} \right)r^{2k} \right\}^{1/2} \\
& \stackrel{(a)}\le  \left\{d\sum_{k=1}^d \left( [C^{(k)}_{i,j}+g_xC^{(k)}_{j,i}]^2 + (1- g^2_x) (C^{(k)}_{j,i})^{2} \right) \right\}^{1/2}
\end{align}
where $(a)$ is due to each term is non-negative, hence $r^2=1$ maximizes the right hand side. It is not hard to see that, to maximize the right hand side, if $\sum_{k=1}^d C^{(k)}_{i,j} C^{(k)}_{j,i}\ge 0$, we take $g_x=1$; otherwise, we take $g_x= \min_x g_x$. Further with $g_0\le \min_x g_x$, we have $g_0\sum_{k=1}^d C^{(k)}_{i,j} C^{(k)}_{j,i}\ge \min_x g_x \sum_{k=1}^d C^{(k)}_{i,j} C^{(k)}_{j,i}$ when $\sum_{k=1}^d C^{(k)}_{i,j} C^{(k)}_{j,i}<  0$.

Therefore, we have the right hand side:
   \begin{align}
& \left\{d\sum_{k=1}^d \left( [C^{(k)}_{i,j}+g_xC^{(k)}_{j,i}]^2 + (1- g^2_x) (C^{(k)}_{j,i})^{2} \right) \right\}^{1/2} \\
&\le
  \sqrt{d}\left\{\max_{h\in \{1,g_0\}}  ( \sum_{k=1}^d
  \left( [C^{(k)}_{i,j}+hC^{(k)}_{j,i}]^2 + (1- h^2) (C^{(k)}_{j,i})^{2} \right)
   \right\}^{1/2}.
   \end{align}

  Therefore, we have
  \begin{align}
&  \sum_{j=1}^{p}|\sum_{k=1}^d [C^{(k)}_{i,j} r^{k}\exp(\mathbb{I}kx)+ C^{(k)}_{j,i} r^{k}\exp(-\mathbb{I}kx)]|\\
&\le \sup_{g_x\in [\min_x g_x,1]} \sum_{j=1}^{p}  \left\{d\sum_{k=1}^d \left( [C^{(k)}_{i,j}+g_xC^{(k)}_{j,i}]^2 + (1- g^2_x) (C^{(k)}_{j,i})^{2} \right) \right\}^{1/2}\\
& \stackrel{(a)}\le \sqrt{d} \sum_{j=1}^{p} \max_{h_j\in \{g_0,1\}} \left\{\sum_{k=1}^d \left( [C^{(k)}_{i,j}+h_{j}C^{(k)}_{j,i}]^2 + (1- h_{j}^{2}) (C^{(k)}_{j,i})^{2} \right) \right\}^{1/2},
  \end{align}
  where $(a)$ is due to the supremum of a sum over $g_x\in  [\min_x g_x,1]$ is smaller or equal to the sum of the supremum of each term, and each supremum is smaller than the one replacing $g_x$ by $h_j\in \{g_0,1\}$.

\subsection{Proof for Theorem 3}
First we note that
$$\Pi\{||c-c_0||>\eta\mid \Phi,y,X\}\le\Pi\{||c-\hat{c}||>\eta/2\mid\Phi y,X\}+\Pi\{||\hat{c}-c_0||>\eta/2\mid \Phi,y,X\},$$ where $\hat{c}=\{\hat{\Gamma}+\Phi^{-1}/N\}^{-1}\hat{\gamma}$.

Let $\Sigma_\varepsilon^{-1}=Q^{\rm T}_\Sigma Q_\Sigma$ be the symmetric decomposition, the Woodbury identity gives:
\(
 &( \Sigma_\varepsilon^{-1}\otimes X^{\rm T}X+\Phi^{-1}) ^{-1} =
 \{  (Q_{\Sigma}^{\rm T}\otimes X^{\rm T}) (Q_{\Sigma}\otimes X)+\Phi^{-1}\} ^{-1} \\
& = \Phi - \Phi   (Q_{\Sigma}^{\rm T}\otimes X^{\rm T}) \{I +   (Q_{\Sigma}\otimes X)\Phi   (Q_{\Sigma}^{\rm T}\otimes X^{\rm T})\}^{-1}  (Q_{\Sigma}^{\rm T}\otimes X^{\rm T})\Phi.
\)
Therefore, if $(i,j)\not\in \bigcup_{l=1}^m T^l$, then with $ \sum_{l=1}^m s_l A^l_{i,j}\to 0$ uniformly, we have $(c_{i,j,k} \mid\ \Sigma_\varepsilon, r,\eta,s,A)$ converge to a point mass at zero. On the other hand, for those $(i,j)\in  \bigcup_{l=1}^m T^l$, due to the lower-boundedness as described in A3, we know $\Phi_{i,j,k}^{-1}< \kappa$ for some constant $\kappa>0$. Moreover, by Assumption A1, for those $(i,j)\notin  \bigcup_{l=1}^m T^l$, we have $\Phi^{-1}_{i,j,k}< \kappa/\epsilon^2$, where $\epsilon$ is defined in A2. Therefore, for any fixed $\epsilon>0$, $||\Phi^{-1}||$ is bounded.

\textbf{(i) Bound the distance between $\hat c$ and $c_0$ given $\Phi$:}
\begin{align}
&\hat{c}-c_0=[\hat{\Gamma}+\Phi^{-1}/N]^{-1}\hat{\gamma}-c_0\\
&=[\hat{\Gamma}+\Phi^{-1}/N]^{-1}(\hat{\gamma}-\hat{\Gamma}c_0+\hat{\Gamma}c_0)-c_0\\
&=[\hat{\Gamma}+\Phi^{-1}/N]^{-1}(\hat{\gamma}-\hat{\Gamma}c_0)+[\hat{\Gamma}+\Phi^{-1}/N]^{-1}(\hat{\Gamma}+\Phi^{-1}/N-\Phi^{-1}/N)c_0-c_0\\
&=[\hat{\Gamma}+\Phi^{-1}/N]^{-1}(\hat{\gamma}-\hat{\Gamma}c_0-\Phi^{-1}/Nc_0).
\end{align}

Hence,
\begin{align}
&||\hat{c}-c_0||\\
&=||[\hat{\Gamma}+\Phi^{-1}/N]^{-1}(\hat{\gamma}-\hat{\Gamma}c_0-\Phi^{-1}c_0/N)||\\
&\leq ||[\hat{\Gamma}+\Phi^{-1}/N]^{-1}||\cdot\{||\Phi^{-1}c_0/N||+||\hat{\gamma}-\hat{\Gamma}c_0||\}.
\label{eq:proof0}
\end{align}

Note that,
\begin{align}
        &||[\hat{\Gamma}+\Phi^{-1}/N]^{-1}||=\lambda_{\max}[\hat{\Gamma}+\Phi^{-1}/N]^{-1}=1/\lambda_{\min}[\hat{\Gamma}+\Phi^{-1}/N]\\
        &\stackrel{(a)}\le1/\{\lambda_{\min}(\hat{\Gamma})+\lambda_{\min}\{\Phi^{-1}/N)\}\le1/\lambda_{\min}(\hat{\Gamma}),
        \end{align}where $(a)$ is due to the positive definiteness of $\Phi^{-1}$ and $\hat{\Gamma}$.
Together with $\lambda_{\min}(\hat{\Gamma})=1/||(\Sigma_\varepsilon^{-1}\otimes X'X/N)^{-1}||=1/(||\Sigma_\varepsilon||\cdot||(X'X/N)^{-1}||)$, we have $||[\hat{\Gamma}+\Phi^{-1}/N]^{-1}||\le ||\Sigma_\varepsilon||/\lambda_{\min}(X'X/N)$.

Thus, together with (\ref{eq:proof0}), we have
\begin{align}
&\Pi(||\hat{c}-c_0||>\eta\mid \Phi,y,X)\\
&\leq \Pi(||[\hat{\Gamma}+\Phi^{-1}/N]^{-1}||\cdot\{||\Phi^{-1}c_0/N||+||\hat{\gamma}-\hat{\Gamma}c_0||\}>\eta\mid \Phi,y,X)\\
&\stackrel{(a)}\leq \Pi(||\Sigma_\varepsilon||/\lambda_{\min}(X'X/N)\cdot (||\Phi^{-1}\cdot c_0/N||+||\hat{\gamma}-\hat{\Gamma}c_0||)> \eta\mid \Phi,y,X)\\
&\le\Pi(\lambda_{\min}(X'X/N)<\lambda_1\mid y,X)+\Pi(||\Phi^{-1}\cdot c_0/N||+||\hat{\gamma}-\hat{\Gamma}c_0||> \eta \lambda_1/||\Sigma_\varepsilon||\mid \Phi,y,X)\\
&\le\Pi(\lambda_{\min}(X'X/N)<\lambda_1\mid y,X)+\Pi(||\Phi^{-1}\cdot c_0/N||>\eta \lambda_1/(2||\Sigma_\varepsilon||)\mid \Phi,y,X)\\
&+\Pi(||\hat{\gamma}-\hat{\Gamma}c_0||>\eta \lambda_1/(2||\Sigma_\varepsilon||)\mid \Phi,y,X),
\label{eq:proof1}
\end{align} where $(a)$ is due to $||[\hat{\Gamma}+\Phi^{-1}/N]^{-1}||\le ||\Sigma_\varepsilon||/\lambda_{\min}(X'X/N)$ and $\lambda_1$ is defined in Proposition B.2 in \cite{Ghosh2019}. Assumption A2 guarantees the validity of Proposition B.2 in \cite{Ghosh2019}, hence the first term on the right hand side of (~\ref{eq:proof1}) is less than $2\exp\{-\sqrt{Np}\}$.

By Assumption A1, for those $(i,j)\in  \bigcup_{l=1}^m T^l$, due to the lower-boundedness as described in A3, we know $\Phi^{-1}_{i,j,k}< \kappa$ for some constant $\kappa>0$. Moreover, for those $(i,j)\notin  \bigcup_{l=1}^m T^l$, we have $\Phi^{-1}_{i,j,k}< \kappa/\epsilon^2$, where $\epsilon$ is defined in A2. Therefore, for any fixed $\epsilon>0$, $||\Phi^{-1}||$ is bounded. Together with Assumption A4 $||c_0||\le K$, we have \begin{align}
    ||\Phi^{-1}c_0||=o(N),
    \label{eq:boundedness}
\end{align}which ensures the second term on the right hand side of (\ref{eq:proof1}) converges to 0 as $N\to\infty$.

Note that $||\hat{\gamma}-\hat{\Gamma}c_0||=||\text{vec}(X^{\rm T}E/N)||=||X^{\rm T}E/N||_F\stackrel{(a)}\le \sqrt{dp}||X^{\rm T}E/N||$, where $(a)$ is due to matrix norm property $||A||_F\le \text{rank}(A) ||A||$. By Corollary B.4 in \cite{Ghosh2019}, $||X^{\rm T}E/N||\le$  $2\pi\lambda_{\max}(\Sigma_\varepsilon)[1+(1+\mu_{\min}(\tilde{C}))/\mu_{\max}(\tilde{C})]\zeta_N$ with probability at least $1-6\exp\{-\sqrt{Np}\}$, where $\zeta_n^2=4{p(d+1)\log21/N+\sqrt{p/N}}/c$.. Therefore, for sufficiently large $N$, we have $||\hat{\gamma}-\hat{\Gamma}c_0||\le \sqrt{dp}2\pi\lambda_{\max}(\Sigma_\varepsilon)[1+(1+\mu_{\min}(\tilde{C}))/\mu_{\max}(\tilde{C})]\zeta_N$ with probability at least $1-6\exp\{-\sqrt{Np}\}$, which implies $\Pi(||\hat{\gamma}-\hat{\Gamma}c_0||>\eta \lambda_1/(2||\Sigma_\varepsilon||)\mid y,X)=0.$ Combining the results above gives
\begin{align}
    &\Pi(||\hat{c}-c_0||>\eta\mid \Phi,y,X)\to0,\text{ as }N\to\infty.
\end{align}

\textbf{(ii) Bound the distance between $\hat c$ and $c$ given $\Phi$:}

Recall that $c\mid \Phi,y,X \sim \text{N}(\hat{c},[\hat{\Gamma}+\Phi^{-1}/N]^{-1}/N)$ and define $\tilde{\Sigma_\varepsilon}$ as $[\hat{\Gamma}+\Phi^{-1}/N]^{-1}/N$. First note that $Z:=\tilde{\Sigma_\varepsilon}^{-\frac{1}{2}}(c-\hat{c})\mid \Phi,y,X \sim \text{N}(\vec{0},I_{dp^2})$.

Also, \begin{align}
&||c-\hat{c}||=||\tilde{\Sigma_\varepsilon}^{1/2} \cdot\tilde{\Sigma_\varepsilon}^{-1/2}(c-\hat{c})|| \\
&=||\tilde{\Sigma_\varepsilon}^{1/2} Z|| \\
&\leq ||\tilde{\Sigma_\varepsilon}^{1/2}|| \cdot ||Z|| \\
&=(N||\tilde{\Sigma_\varepsilon}||)^{1/2} \cdot ||Z/\sqrt{N}||.
\end{align}

As previously stated, we have, \begin{align}
&N||\tilde{\Sigma_\varepsilon}||= ||[\hat{\Gamma}+\Phi^{-1}/N]^{-1}||\leq ||\Sigma_\varepsilon||/\lambda_{\min}(X'X/N).
\end{align}

Therefore,
\begin{align}
&||c-\hat{c}||\leq \{||\Sigma_\varepsilon||/\lambda_{\min}(X'X/N)\}^{1/2} \cdot||Z/\sqrt{N}||.
\end{align}

Hence,
\begin{align}
&\Pi(||c-\hat{c}||>\eta\mid \Phi,y,X)\\
&\leq \Pi(\{||\Sigma_\varepsilon||/\lambda_{\min}(X'X/N)\}^{1/2} \cdot||Z/\sqrt{N}||>\eta\mid \Phi,y,X) \\
&\leq \Pi(\lambda_{\min}(X'X/N)< \lambda_1) + \Pi(||\Sigma_\varepsilon||^{1/2}||Z/\sqrt{N}||>\eta(\lambda_1)^{1/2} \mid \Phi,y,X) \\
&=\Pi(\lambda_{\min}(X'X/N)< \lambda_1)+ \Pi( ||Z/\sqrt{N}||>\eta(\lambda_1/||\Sigma_\varepsilon||)^{1/2} \mid \Phi,y,X)\\
&=\Pi(\lambda_{\min}(X'X/N)< \lambda_1)+ \Pi( ||Z||^2>N\eta^2\lambda_1/||\Sigma_\varepsilon|| \mid \Phi,y,X). \label{eq:proof2}
\end{align}
By Proposition B.2 in \cite{Ghosh2019}, the first term on the right hand side of (\ref{eq:proof2}) converges to 0 as $N\to\infty$.

Next, we show $\Pi( ||Z||^2>N\eta^2\lambda_1/||\Sigma_\varepsilon \mid \Phi,y,X)\to0$, as $N\to\infty$.Note that $||Z||^2\sim \chi_2(dp^2)$, which implies $\mathbb{E}_0(||Z||^2)=dp^2$, $\mathbb{V}_0(||Z||^2)=2dp^2$. Using Chebyshev's inequality gives
\begin{align}
    &\Pi\{ |||Z||^2-dp^2|>N\eta^2\lambda_1/||\Sigma_\varepsilon|| \mid \Phi,y,X\}<2dp^2/(N\eta^2\lambda_1/||\Sigma_\varepsilon||)^2\\
    &=2(p/N)^2\cdot d||\Sigma_\varepsilon||^2/(\eta^2\lambda_1).
\end{align}

Thus,
\begin{align}
    &\Pi( ||Z||^2>N\eta^2\lambda_1/||\Sigma_\varepsilon \mid \Phi,y,X)\\
    &\le \Pi( |||Z||^2-dp^2|>N\eta^2\lambda_1/||\Sigma_\varepsilon \mid \Phi,y,X)+\Pi( dp^2>N\eta^2\lambda_1/||\Sigma_\varepsilon \mid \Phi,y,X).
\end{align}

Combining the two sections above, we have
\begin{align}
    &\Pi\{||c-c_0||>\eta\mid \Phi,y,X\}\to0, \ t{ as }N\to\infty.
    \end{align}

\textbf{(iv) Show convergence of posterior probability}

 First note that $c\mid\Phi,y,X\sim \text{N}(\hat{c},\frac{1}{N}(\hat{\Gamma}+\frac{\Phi^{-1}}{N})^{-1})$.

 We obtain the marginal posterior distribution of $(\Phi\mid y,X)$ by integrating out the regression coefficient $c$:
\begin{align}
    &\Pi(\Phi\mid y,X)\propto  \int \mathcal{L}(c,\Phi;y)\pi_0(c|\Phi)\pi_0(\Phi) \,\textup{d} c \\
   &\propto \int \exp\{-\frac{1}{2}\{y-(I_p\otimes X)c\}^{\rm T}(\Sigma_\varepsilon^{-1} \otimes I_{T-d}) \{y-(I_p\otimes X)c\}\}\exp\{-\frac{1}{2}c^{\rm T}\Phi^{-1}c\}\\
   &\qquad \{\det(\Phi)\}^{-1/2}\pi_0(\Phi)\,\textup{d} c \\
    &\propto \{ \det(\hat{\Gamma}+\Phi^{-1}/N)\}^{-1/2}\pi_0(\Phi)\{\det(\Phi)\}^{-1/2}\exp\{N\hat{\gamma}^{\rm
T}[\hat{\Gamma}+\Phi^{-1}/N]^{-1}\hat{\gamma}/2\}.
\end{align}

Comparing the posterior densities of two $\Phi_{\star\star}$ and $\Phi_{\star}$:
\begin{align}
    \frac{\Pi(\Phi_{\star\star}\mid y,X)}{\Pi(\Phi_{\star}\mid y,X)}
   &=\frac{\det^{-1/2} [\hat{\Gamma}+\Phi_{\star\star}^{-1}/N]\det^{-1/2}(\Phi_{\star\star})\pi_0(\Phi_{\star\star})}{\det^{-1/2} [\hat{\Gamma}+\Phi_{\star}^{-1}/N]\det^{-1/2}(\Phi_{\star})\pi_0(\Phi_{\star})} \\ & \times \exp\left[N\hat{\gamma}^{\rm T}\left\{[\hat{\Gamma}+\Phi_{\star\star}^{-1}/N]^{-1} -[\hat{\Gamma}+\Phi_{\star}^{-1}/N]^{-1}\right\}   \hat{\gamma}/2 \right ].
\end{align}

By Assumption A1, we know $\Phi_{i,j,k,\star\star}^{-1}\le\min\{\kappa,\kappa/\epsilon^2\}$. Since $\Phi_{\star\star}^{-1}$ is diagonal, this guarantees the boundedness of $\text{det}\{\Phi_{\star\star}^{-1}\}$. Similarly, $\text{det}\{\Phi_\star^{-1}\}$ is bounded.

Next, we show $\det\{\hat{\Gamma}+\Phi_{\star\star}^{-1}/N\}$ is bounded for sufficiently large $N$. Then together with $\text{det}^{\frac{1}{2}}(\Gamma+\Phi_\star^{-1}/N)$ is greater than 0, we obtain the boundedness of $\text{det}^{\frac{1}{2}}(\Gamma+\Phi_{\star\star}^{-1}/N)$ and $\text{det}^{\frac{1}{2}}(\Gamma+\Phi_\star^{-1}/N)$ for sufficiently large $N$.

By Assumption A2 $0<\sup_{N\geq 1}\lambda_{\max}(\Gamma_y(0))<\infty$ and  Proposition B.2 in \cite{Ghosh2019}, there exists $0<\lambda_2<\infty$, such that $\Pi\{\lambda_{\max}(X'X/N)>\lambda_2\}\le2\exp\{-2\sqrt{Np}\}$. This ensures that $\lambda_{\max}(X'X/N)<\lambda_2$ with probability at least $1-2\exp\{-2\sqrt{Np}\}$. Also note that $\text{det}\{\hat{\Gamma}+\Phi_{\star\star}^{-1}/N\}=\prod\limits_{i=1}^{dp^2}\lambda_{i}\{\hat{\Gamma}+\Phi_{\star\star}^{-1}/N\}$, where $\lambda_{i}\{\hat{\Gamma}+\Phi_{\star\star}^{-1}/N\}$ is the $i^{\text{th}}$ largest eigenvalue of $\hat{\Gamma}+\Phi{\star\star}^{-1}/N$.

$\lambda_{\max}(\hat{\Gamma}+\Phi_{\star\star}^{-1}/N)\le\lambda_{\max}(\hat{\Gamma})+\lambda_{\max}(\Phi{\star\star}^{-1}/N)$ leads to $\text{det}\{\hat{\Gamma}+\Phi_{\star\star}^{-1}/N\}\le[\lambda_{\max}(\hat{\Gamma})+\lambda_{\max}(\Phi{\star\star}^{-1}/N)]^{dp^2}$. Also, $\lambda_{\max}(\hat{\Gamma})=||\Sigma_\varepsilon||\lambda_{\max}(X'X/N)\le\lambda_2||\Sigma_\varepsilon||$ with probability at least $1-2\exp\{-2\sqrt{Np}\}$, together with $\lambda_{\max}(\{\Phi_{\star\star}^{-1}/N)$ is bounded as $N$ goes to $\infty$, we have $\text{det}\{\hat{\Gamma}+\Phi_{\star\star}^{-1}/N\}$ is bounded with probability at least $1-2\exp\{-2\sqrt{Np}\}$.

Without loss of generality, we only consider $\Phi_\star$ with $\pi_0(\Phi_{\star\star})$ is greater than 0, which guarantees the boundedness of the term $\pi_0(\Phi_{\star\star})/\pi_0(\Phi_\star)$. Combining the results above, for sufficiently large $N$, we have
\begin{align}
    &\frac{\det^{-1/2} [\hat{\Gamma}+\Phi_{\star\star}^{-1}/N]\det^{-1/2}(\Phi_{\star\star})\pi_0(\Phi_{\star\star})}{\det^{-1/2} [\hat{\Gamma}+\Phi_{\star}^{-1}/N]\det^{-1/2}(\Phi_{\star})\pi_0(\Phi_{\star})}\le H(X,\tau),
    \label{eq:proof10}
\end{align}where $H(X,\tau)$ is a bounded positive function.

Next, we focus on $\exp\{N/2\hat{\gamma}^{\rm T}(\hat{\Gamma}+\Phi_{\star\star}^{-1}/N)^{-1}\hat{\gamma}-\frac{N}{2}\hat{\gamma}^{\rm T}(\hat{\Gamma}+\Phi_{\star}^{-1}/N)^{-1}\hat{\gamma}\}$ and show it goes to 0 in probability, with fixed $p$ and sufficiently large $N$.

We divide $N\hat{\gamma}^{\rm T}(\hat{\Gamma}+\frac{\Phi_{\star\star}^{-1}}{N})^{-1}\hat{\gamma}$ into three parts:
\begin{align}
    &N\hat{\gamma}^{\rm T}[\hat{\Gamma}+\Phi_{\star\star}^{-1}/N]^{-1}\hat{\gamma}\\
   &=N(\hat{\gamma}-\hat{\Gamma}c_0+\hat{\Gamma}c_0)^{\rm T}[\hat{\Gamma}+\Phi_{\star\star}^{-1}/N]^{-1}(\hat{\gamma}-\hat{\Gamma}c_0+\hat{\Gamma}c_0)\\
    &=N(\hat{\Gamma}c_0)^{\rm T}[\hat{\Gamma}+\Phi_{\star\star}^{-1}/N]^{-1}\hat{\Gamma}c_0\\
    &+N(\hat{\gamma}-\hat{\Gamma}c_0)^{\rm T}[\hat{\Gamma}+\Phi_{\star\star}^{-1}/N]^{-1}(\hat{\gamma}-\hat{\Gamma}c_0)\\
    &+2N(\hat{\Gamma}c_0)^{\rm T}[\hat{\Gamma}+\Phi_{\star\star}^{-1}/N]^{-1}(\hat{\gamma}-\hat{\Gamma}c_0).
    \end{align}

Thus,
\begin{align}
    &N\hat{\gamma}^{\rm T}[\hat{\Gamma}+\Phi_{\star\star}^{-1}/N]^{-1}\hat{\gamma}-N\hat{\gamma}^{\rm T}[\hat{\Gamma}+\Phi_{\star}^{-1}/N]^{-1}\hat{\gamma}\\
    &=\underbrace{\{N(\hat{\Gamma}c_0)^{\rm T}[\hat{\Gamma}+\Phi_{\star\star}^{-1}/N]^{-1}\hat{\Gamma}c_0-N(\hat{\Gamma}c_0)^{\rm T}[\hat{\Gamma}+\Phi_{\star}^{-1}/N]^{-1}\hat{\Gamma}c_0\}}_{\mathcal{I}_1}\\
    &+\underbrace{N(\hat{\gamma}-\hat{\Gamma}c_0)^{\rm T}\{[\hat{\Gamma}+\Phi_{\star\star}^{-1}/N]^{-1}-[\hat{\Gamma}+\Phi_{\star}^{-1}/N]^{-1}\}(\hat{\gamma}-\hat{\Gamma}c_0)}_{\mathcal{I}_2}\\
    &+\underbrace{2N(\hat{\Gamma}c_0)^{\rm T}\{[\hat{\Gamma}+\Phi_{\star\star}^{-1}/N]^{-1}-[\hat{\Gamma}+\Phi_{\star}^{-1}/N]^{-1}\}(\hat{\gamma}-\hat{\Gamma}c_0)}_{\mathcal{I}_3}.
    \label{eq:proof3}
\end{align}

Next, we show $\mathcal{I}_1$ approaches negative infinity and $\mathcal{I}_2$, $\mathcal{I}_3$ are bounded for sufficiently small $\tau$ and sufficiently large $N$.

Consider $\mathcal{I}_1$ from (\ref{eq:proof3}),
\begin{align}
    &N(\hat{\Gamma}c_0)^{\rm T}[\hat{\Gamma}+\Phi_{\star\star}^{-1}/N]^{-1}\hat{\Gamma}c_0\\
    &=N(c_0)^{\rm T}\hat{\Gamma}[\hat{\Gamma}+\Phi_{\star\star}^{-1}/N]^{-1}\hat{\Gamma}c_0\\
    &=N(c_0)^{\rm T}(\hat{\Gamma}+\Phi_{\star\star}^{-1}/N-\Phi^{-1}/N)[\hat{\Gamma}+\Phi_{\star\star}^{-1}/N]^{-1}\hat{\Gamma}c_0\\
    &=N(c_0)^{\rm T}\hat{\Gamma}c_0-N(c_0)^{\rm T}(\Phi^{-1}/N)[\hat{\Gamma}+\Phi_{\star\star}^{-1}/N]^{-1}\hat{\Gamma}c_0\\
    &=N(c_0)^{\rm T}\hat{\Gamma}c_0-N(c_0)^{\rm T}(\Phi_{\star\star}^{-1}/N)[\hat{\Gamma}+\Phi_{\star\star}^{-1}/N]^{-1}(\hat{\Gamma}+\Phi_{\star\star}^{-1}/N-\Phi_{\star\star}^{-1}/N)c_0\\
    &=N(c_0)^{\rm T}\hat{\Gamma}c_0-(c_0)^{\rm T}\Phi_{\star\star}^{-1}c_0+N(c_0)^{\rm T}(\Phi_{\star\star}^{-1}/N)[\hat{\Gamma}+\Phi_{\star\star}^{-1}/N]^{-1}(\Phi_{\star\star}^{-1}/N)c_0.
\end{align}

Thus, $\mathcal{I}_1$ is equivalent to
\begin{align}
    &N(\hat{\Gamma}c_0)^{\rm T}[\hat{\Gamma}+\Phi_{\star\star}^{-1}/N]^{-1}\hat{\Gamma}c_0-N(\hat{\Gamma}c_0)^{\rm T}[\hat{\Gamma}+\Phi_{\star}^{-1}/N]^{-1}\hat{\Gamma}c_0\\
    &=(c_0)^{\rm T}\Phi^{-1}_{\star\star}c_0-N(c_0)^{\rm T}(\Phi_{\star\star}^{-1}/N)[\hat{\Gamma}+\Phi_{\star\star}^{-1}/N]^{-1}(\Phi_{\star\star}^{-1}/N)c_0\\
    &-(c_0)^{\rm T}\Phi^{-1}_{\star}c_0+N(c_0)^{\rm T}(\Phi_{\star}^{-1}/N)[\hat{\Gamma}+\Phi_{\star}^{-1}/N]^{-1}(\Phi_{\star}^{-1}/N)c_0.
    \label{eq:proof4}
\end{align}

Consider the first term on the right hand side of (\ref{eq:proof4}). Since $G_0\subseteq \cup_{l=1}^{m}T_{\star}^l$, we have $c_{0,i,j,k}=0$ for $(i,j)\notin \cup_{l=1}^{m}T_{\star}^l$. If $(i,j)\in \cup_{l=1}^{m}T_{\star}^l$, $\Phi_{(i,j,k)}^{-1}\le\kappa$. This implies  $(c_0)^{\rm T}\Phi^{-1}_{\star}c_0\le \kappa ||c_0||^2$.

Also, we show the third term on the right hand side of (\ref{eq:proof4}) approaches infinity as $\tau$ goes to 0.Also note that $G_0\nsubseteq \cup_{l=1}^{m}T^l$. This implies that there exists some $(i,j)\notin \cup_{l=1}^{m}T_{\star\star}^l$ such that $c_{0,i,j,k}>\zeta$, where $\zeta$ is a positive constant. Moreover,$ \sum_{l=1}^m s_l A^l_{i,j}\to 0$ uniformly, which ensures  $(c_0)^{\rm T}\Phi_{\star\star}^{-1}c_0 \to \infty$ as $\tau \to 0$. Hence, there exists $M_\tau>0$ such that $(c_0)^{\rm T}\Phi^{-1}c_0\ge M_\tau$ and $M_\tau\to\infty$, as $\tau\to0$.

Note that,
\begin{align}
    &N(c_0)^{\rm T}(\Phi_{\star\star}^{-1}/N)[\hat{\Gamma}+\Phi_{\star\star}^{-1}/N]^{-1}(\Phi_{\star\star}^{-1}/N)c_0 \\
    &\le ||c_0||^2 ||\Phi_{\star\star}^{-1}[\hat{\Gamma}+\Phi_{\star\star}^{-1}/N]^{-1}\Phi_{\star\star}^{-1}||/N\\
    &\le ||c_0||^2 ||\Phi_{\star\star}^{-1}||^2||[\hat{\Gamma}+\Phi_{\star\star}^{-1}/N]^{-1}||/N\\
    &\stackrel{(a)}= ||c_0||^2 ||\Phi_{\star\star}^{-1}||_\infty^2||[\hat{\Gamma}+\Phi_{\star\star}^{-1}/N]^{-1}||/N.
\end{align} where $(a)$ is due to $\Phi_{\star\star}$ is diagonal and positive definite, hence its largest eigenvalue is the maximal value on the diagonal.

Next, we obtain an upper bound of the second term on the right hand side of (\ref{eq:proof4}). Using $||[\hat{\Gamma}+\Phi^{-1}/N]^{-1}||\le ||\Sigma_\varepsilon||/\lambda_{\min}(X'X/N)$ gives
\begin{align}
    &N(c_0)^{\rm T}(\Phi_{\star\star}^{-1}/N)[\hat{\Gamma}+\Phi_{\star\star}^{-1}/N]^{-1}(\Phi_{\star\star}^{-1}/N)c_0\\
    &\le ||\Sigma_\varepsilon||/\lambda_{\min}(X'X/N)\cdot||\Phi_{\star\star}^{-1}c_0||^2/N\\
    &\le||\Sigma_\varepsilon||/\lambda_{\min}(X'X/N)||c_0||^2||\Phi_{\star\star}^{-1}||_\infty^2/N,
\end{align} which converges to 0 as $N\to\infty$.

Similarly, for the fourth term on the right-hand side of (~\ref{eq:proof4}), \begin{align}
    &N(c_0)^{\rm T}(\Phi_{\star}^{-1}/N)[\hat{\Gamma}+\Phi_{\star}^{-1}/N]^{-1}(\Phi_{\star}^{-1}/N)c_0\le||\Sigma_\varepsilon||/\lambda_{\min}(X'X/N)||c_0||^2||\Phi_{\star}^{-1}||_\infty^2/N.
\end{align}
Thus, \begin{align}
    &\mathcal{I}_1\le -M_\tau+\kappa||c_0||^2+||\Sigma_\varepsilon||/\lambda_{\min}(X'X/N)||c_0||^2(||\Phi_{\star}^{-1}||_\infty^2+||\Phi_{\star\star}^{-1}||_\infty^2)/N.
    \label{eq:proof5}
\end{align}

Since $||\Phi^{-1}||_\infty\le \min\{\kappa/\epsilon^2,\kappa\}$ and $||\Phi_\star^{-1}||_\infty\le \min\{\kappa/\epsilon^2,\kappa\}$, together with (\ref{eq:proof5}), we have $\mathcal{I}_1$ approach $-\infty$ with sufficiently small $\tau>0$ and sufficiently large $N=N(\tau)$.

Next, we show $\mathcal{I}_2=N(\hat{\gamma}-\hat{\Gamma}c_0)^{\rm T}\{[\hat{\Gamma}+\Phi_{\star\star}^{-1}/N]^{-1}-[\hat{\Gamma}+\Phi_{\star}^{-1}/N]^{-1}\}(\hat{\gamma}-\hat{\Gamma}c_0)$ is bounded above.

By Corollary B.4 in \cite{Ghosh2019}, we have $||\hat{\gamma}-\hat{\Gamma}c_0||\le\mathbb{Q}(c_0,\Sigma_\varepsilon)\sqrt{dp^2/N}$ with probability at least $1-6\exp\{-\sqrt{Np}\}$, where $\mathbb{Q}(c_0,\Sigma_\varepsilon)=2\pi\lambda_{\max}(\Sigma_\varepsilon)[1+(1+\mu_{\min}(\tilde{C}))/\mu_{\max}(\tilde{C})]\zeta_N$, where where $\zeta_n^2=4{p(d+1)\log21/N+\sqrt{p/N}}/h$.\\ Together with  $||[\hat{\Gamma}+\Phi^{-1}/N]^{-1}||\le ||\Sigma_\varepsilon||/\lambda_{\min}(X'X/N)$, we have
\begin{align}
    &N(\hat{\gamma}-\hat{\Gamma}c_0)^{\rm T}[\hat{\Gamma}+\Phi_{\star\star}^{-1}/N]^{-1}(\hat{\gamma}-\hat{\Gamma}c_0))\\
    &\le N||[\hat{\Gamma}+\Phi_{\star\star}^{-1}/N]^{-1}||\cdot||\hat{\gamma}-\hat{\Gamma}c_0||^2\\
    &\le ||\Sigma_\varepsilon||/\lambda_{\min}(X'X/N) \mathbb{Q}(c_0,\Sigma_\varepsilon)^2 dp^2/N .
\end{align}

Similarly, we obtain
\begin{align}
    &N(\hat{\gamma}-\hat{\Gamma}c_0)^{\rm T}[\hat{\Gamma}+\Phi_{\star}^{-1}/N]^{-1}(\hat{\gamma}-\hat{\Gamma}c_0))\le||\Sigma_\varepsilon||/\lambda_{\min}(X'X/N) \mathbb{Q}(c_0,\Sigma_\varepsilon)^2 dp^2/N.
\end{align}
Hence,
\begin{align}
    \mathcal{I}_2\le2||\Sigma_\varepsilon||/\lambda_{\min}(X'X/N)  \mathbb{Q}(c_0,\Sigma_\varepsilon)^2 dp^2/N.
\end{align}

Finally, we show $\mathcal{I}_3=2N(\hat{\Gamma}c_0)^{\rm T}\{[\hat{\Gamma}+\Phi_{\star\star}^{-1}/N]^{-1}-[\hat{\Gamma}+\Phi_{\star}^{-1}/N]^{-1}\}(\hat{\gamma}-\hat{\Gamma}c_0)$ is bounded, as $N\to\infty$.

Note that
\begin{align}
    &N(\hat{\Gamma}c_0)^{\rm T}[\hat{\Gamma}+\Phi_{\star\star}^{-1}/N]^{-1}(\hat{\gamma}-\hat{\Gamma}c_0)\\
    &=N(c_0)^{\rm T}[\hat{\Gamma}+\Phi_{\star\star}^{-1}/N-\Phi_{\star\star}^{-1}/N][\hat{\Gamma}+\Phi_{\star\star}^{-1}/N]^{-1}(\hat{\gamma}-\hat{\Gamma}c_0)\\
    &=N(c_0)^{\rm T}(\hat{\gamma}-\hat{\Gamma}c_0)-(c_0)^{\rm T}\Phi_{\star\star}^{-1}[\hat{\Gamma}+\Phi_{\star\star}^{-1}/N]^{-1}(\hat{\gamma}-\hat{\Gamma}c_0).
\end{align}

Similarly, we have \begin{align}
    &N(\hat{\Gamma}c_0)^{\rm T}[\hat{\Gamma}+\Phi_{\star}^{-1}/N]^{-1}(\hat{\gamma}-\hat{\Gamma}c_0)\\
    &=N(c_0)^{\rm T}(\hat{\gamma}-\hat{\Gamma}c_0)-(c_0)^{\rm T}\Phi_{\star\star}^{-1}[\hat{\Gamma}+\Phi_{\star}^{-1}/N]^{-1}(\hat{\gamma}-\hat{\Gamma}c_0).
\end{align}

Thus,
\begin{align}
    &\mathcal{I}_3=(c_0)^{\rm T}\Phi_{\star\star}^{-1}[\hat{\Gamma}+\Phi_{\star\star}^{-1}/N]^{-1}(\hat{\gamma}-\hat{\Gamma}c_0)-(c_0)^{\rm T}\Phi_{\star}^{-1}[\hat{\Gamma}+\Phi_{\star}^{-1}/N]^{-1}(\hat{\gamma}-\hat{\Gamma}c_0).
    \label{eq:proof6}
\end{align}

Considering the first term on the right hand side of (\ref{eq:proof6}), we have
\begin{align}
    &|(c_0)^{\rm T}\Phi_{\star\star}^{-1}[\hat{\Gamma}+\Phi_{\star\star}^{-1}/N]^{-1}(\hat{\gamma}-\hat{\Gamma}c_0)|\\
    &\le||\Phi_{\star\star}^{-1}||\cdot ||[\hat{\Gamma}+\Phi_{\star\star}^{-1}/N]^{-1}||\cdot ||c_0||\cdot ||\hat{\gamma}-\hat{\Gamma}c_0||.
    \label{eq:proof7}
\end{align}

Note that $||\Phi_{\star\star}^{-1}||$ is bounded and $||[\hat{\Gamma}+\Phi_{\star\star}^{-1}/N]^{-1}||\le\Sigma_\varepsilon||/\lambda_{\min}(X'X/N)$.

Continuing the inequality of (\ref{eq:proof7}), we have
\begin{align}
    &|(c_0)^{\rm T}\Phi_{\star\star}^{-1}[\hat{\Gamma}+\Phi_{\star\star}^{-1}/N]^{-1}(\hat{\gamma}-\hat{\Gamma}c_0)|\\
    &\le ||c_0||\cdot ||\hat{\gamma}-\hat{\Gamma}c_0|| \Sigma_\varepsilon||/\lambda_{\min}(X'X/N).
\end{align}

Similarly for $\Phi_\star$, the following holds, \begin{align}
    &|(c_0)^{\rm T}\Phi_{\star}^{-1}[\hat{\Gamma}+\Phi_{\star}^{-1}/N]^{-1}(\hat{\gamma}-\hat{\Gamma}c_0)|\\
    &\le ||c_0||\cdot ||\hat{\gamma}-\hat{\Gamma}c_0|| \Sigma_\varepsilon||/\lambda_{\min}(X'X/N).
\end{align}

Therefore,
\begin{align}
    &\mathcal{I}_3\le 2||c_0||\cdot ||\hat{\gamma}-\hat{\Gamma}c_0|| \Sigma_\varepsilon||/\lambda_{\min}(X'X/N).
\end{align}

Combining the results about $\mathcal{I}_1$, $\mathcal{I}_2$ and $\mathcal{I}_3$, with sufficiently large $N=N(\tau)$, the following holds,
\begin{align}
    &\mathcal{I}_1+\mathcal{I}_2+\mathcal{I}_3
    \le -M_\tau+\kappa||c_0||^2+||\Sigma_\varepsilon||/\lambda_{\min}(X'X/N)||c_0||^2(||\Phi_{\star\star}^{-1}||_\infty^2+||\Phi_{\star}^{-1}||_\infty^2)/N\\
    &+2||\Sigma_\varepsilon||/\lambda_{\min}(X'X/N)  \mathbb{Q}(c_0,\Sigma_\varepsilon)^2 dp^2/N\\
    &+2||c_0||\cdot ||\hat{\gamma}-\hat{\Gamma}c_0|| \Sigma_\varepsilon||/\lambda_{\min}(X'X/N).
\end{align}

By Assumption 2, we know $||\Phi^{-1}||_\infty$ and $||\Phi_\star^{-1}||_\infty$ are less than $\min\{\kappa/\epsilon^2,\kappa\}$. Note that Assumption A4 and A5 ensure the boundedness of $||c_0||_\infty$ and $||\Sigma_\varepsilon||$, respectively.

There exists sufficiently large $N_1(\tau)$, such that for $N>N_1(\tau)$,
\begin{align}
    &||\Sigma_\varepsilon||\cdot||c_0||^2(||\Phi_{\star\star}^{-1}||_\infty^2+||\Phi_\star^{-1}||_\infty^2)/N\le 1/2.
    \label{eq:proof8}
\end{align}

For $p=o(N^{1/2})$, $||\hat{\gamma}-\hat{\Gamma}c_0||=o(1)$, the explanation is as follows. since $||\hat{\gamma}-\hat{\Gamma}c_0||=||\text{vec}(X^{\rm T}E/N)||$ and $||\text{vec}(X^{\rm T}E/N)||=||X^{\rm T}E/N||_F\le\sqrt{dp}||X^{\rm T}E/N||$, if $p=o(N^{1/2})$, by Corollary B.4 in \cite{Ghosh2019}, we have $||\hat{\gamma}-\hat{\Gamma}c_0||\to0, \text{ as }N\to\infty$.

Similarly, for $\tau$ given above, there exists sufficiently large $N_2(\tau)$, such that for $N>N_2(\tau)$,
\begin{align}
   &2||c_0||\cdot ||\hat{\gamma}-\hat{\Gamma}c_0|| \Sigma_\varepsilon||/\lambda_{\min}(X'X/N)\le 1/2.
   \label{eq:proof9}
\end{align}

Let $N^\star=\max\{N_1(\tau),N_2(\tau)\}$. Combining (\ref{eq:proof8}) and (\ref{eq:proof9}), for $N>N^\star$, the following holds,
\begin{align}
    &\mathcal{I}_1+\mathcal{I}_2+\mathcal{I}_3\le -M_\tau+\kappa||c_0||^2+\{2||\Sigma_\varepsilon||  \mathbb{Q}(c_0,\Sigma_\varepsilon)^2 dp^2 /N+1\}/\lambda_{\min}(X'X/N).
\end{align}

Thus,  \begin{align}
    &\frac{\Pi(\Phi_{\star\star}\mid y,X)}{\Pi(\Phi_{\star}\mid y,X)}\\
    &\le H(X,\tau)\exp\{-M_\tau+\kappa||c_0||^2+\{2||\Sigma_\varepsilon||  \mathbb{Q}(c_0,\Sigma_\varepsilon)^2dp^2 /N+1\}/\lambda_{\min}(X'X/N)\},
\end{align} where $H(X,\tau)$ is the bounded function in (\ref{eq:proof10}).

For any $\eta>0$, we have,
\begin{align}
    &\Pi(\frac{\Pi(\Phi_{\star\star}\mid y,X)}{\Pi(\Phi_{\star}\mid y,X)}>\eta)\\
    &\le\Pi(\lambda_{\min}(X'X/N)<\lambda_1)+\Pi(\frac{\Pi(\Phi_{\star\star}\mid y,X)}{\Pi(\Phi_{\star}\mid y,X)}>\eta,\lambda_{\min}(X'X/N)>\lambda_1)\\
    &\stackrel{(a)}\le\Pi(\lambda_{\min}(X'X/N)<\lambda_1)\\
    &+\Pi(H(X,\tau)\exp\{-M_\tau+\kappa||c_0||^2+\{2||\Sigma_\varepsilon||  \mathbb{Q}(c_0,\Sigma_\varepsilon)^2dp^2 /N+1\}/\lambda_1\}>\eta),
    \label{eq:proof11}
\end{align}where $(a)$ is due to (\ref{eq:proof10}) and the condition $\lambda_{\min}(X'X/N)>\lambda_1$.

By Proposition 1, the first term on the right hand side of (\ref{eq:proof11}) is less than $2\exp\{-\sqrt{Np}\}$. Consider the second term on the right hand side of (\ref{eq:proof11}).

 First note that $\kappa||c_0||^2+\{2||\Sigma_\varepsilon||  \mathbb{Q}(c_0,\Sigma_\varepsilon)^2 dp^2 /N+1\}/\lambda_1$ and $H(X,\tau)$ are bounded given $c_0$, $\Sigma_\varepsilon$ and $\tau$. Also, we know $M_\tau\to\infty$, as $\tau\to0$.  Therefore for $\eta$ given above, there exists sufficiently small $\tau=\tau(\eta)$ such that
\begin{align}
    &H(X,\tau)\exp\{-M_\tau+\kappa||c_0||^2+\{2||\Sigma_\varepsilon||  \mathbb{Q}(c_0,\Sigma_\varepsilon)^2 dp^2 /N+1\}/\lambda_1\}\le\eta.
\end{align} which guarantees $\Pi(\frac{\Pi(\Phi\mid y,X)}{\Pi(\Phi_{\star}\mid y,X)}>\eta,\lambda_{\min}(\frac{X'X}{N})>\lambda_1)=0$.

Hence, for sufficiently small $\tau=\tau(\eta)$ and $N>N^\star(\tau,\eta)$, \begin{align}
    &\Pi(\frac{\Pi(\Phi\mid y,X)}{\Pi(\Phi_{\star}\mid y,X)}>\eta)\le 2\exp\{-\sqrt{Np}\}.
\end{align}

\section{Calculation of the Tree Rank}
We have been focusing on {\em regularizing} the graph estimates using the tree rank. On the other hand, when given an undirected graph $\bar G$, one may be interested in directly {\em calculating} the tree rank. This is not only useful for properly setting up our simulations later, but also of independent interests. Therefore, we briefly review the relevant results and provide a simplified algorithm.

For a given covering $\bigcup_{l=1}^m T^l\supseteq \bar G$, we can remove some edges from each tree, starting from deleting edges not found in $\bar G$, $F^1= T^1 \cap \bar G$, then sequentially for $l=2,\ldots,m$, removing edges previously covered, $F^l= \{T^l \setminus (\bigcup_{h=1}^{l-1} F^h) \} \cap \bar G$. Each obtained graph $F^l$ is known as a ``forest'', an acyclic graph with possible disconnectivity. It is not hard to see that $\bar G = \cup_{l=1}^m F^l$, $F^l\cap F^h=\varnothing$ for any $l\neq h$; further, the tree rank is exactly equal to the minimum covering number using forests. Using
\(
\text{Tree-Rank}(\bar G)= \max_{H\subseteq \bar G} \left\lceil \frac{|E_H|}{|V_H|-1} \right\rceil,
\)
 we can maximize over all subgraphs of $\bar G$ which has cardinality of $O(2^p)$, efficient search algorithm such as \cite{gabow1992forests} has been developed. Briefly speaking, their algorithm is a combination of solving $k$-forest problems (covering as many edges in $\bar G$ as possible using $k$ forests) and a binary search for the minimum $k$ that covers all the edges in $\bar G$. Due to the high complexity, we refer the readers to that article for the details. In the meantime, we present an approximate algorithm that is much easier to implement.

Let $W^1$ be a $p\times p$ weight matrix with $W^1_{i,j}=1$ if $(i,j)\in \bar G$, and $W^1_{i,j}=0$ otherwise.
Let $W^1$ be a $p\times p$ weight matrix with $W^1_{i,j}=1$ if $(i,j)\in \bar G$, and $W^1_{i,j}=0$ otherwise.

\begin{algorithm}[H]
\SetAlgoLined
 \While{$W^l\neq O$}{
        Find the maximum spanning tree of a complete graph with weight matrix $W^l$, denote the produced tree by $T^l$ and its adjacency matrix by $A_{T^l}$\;
        Set $W^{l+1}=W^l \circ (J-A_{T^l})$\;
        Set $l\leftarrow l+1$\;
 }
 Set $m=l-1$.
 \caption{Find an upper bound estimate $m\ge$Tree-Rank($\bar G$).}
\end{algorithm}

\noindent In the above, $O$ denotes the $p\times p$ matrix filled by zeros, and $J$ the matrix by ones; and one can use Prim's algorithm to easily find the maximum spanning tree \citep{prim1957shortest}.

\section{Stability of the Estimated Autogressive Process in the Data Application}

For the data application, we plot the spectral norm of each companion-form matrix [as the coefficient matrix in the VAR($1$) equivalent representation for the VAR($d$) model] associated with each sampled $\bar C$, all of them are strictly smaller than $1$, which shows the stability of the process.

 \begin{figure}[H]
  \centering
 \subfloat[Proposed model.]
  {\includegraphics[width=.22\textwidth]{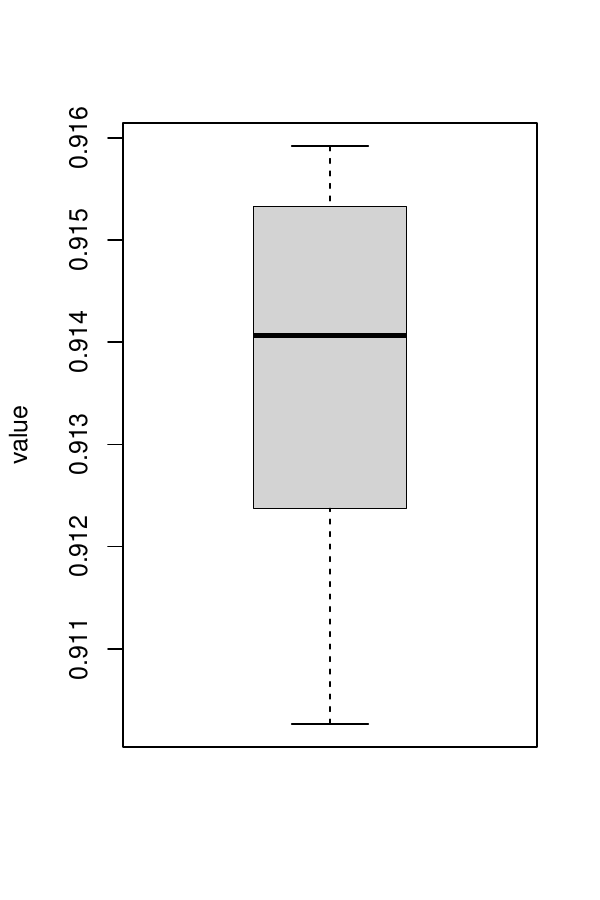}}\hfil
   \subfloat[Element-wise edge selection alone (using generalized Pareto).]
  {\includegraphics[width=.22\textwidth]{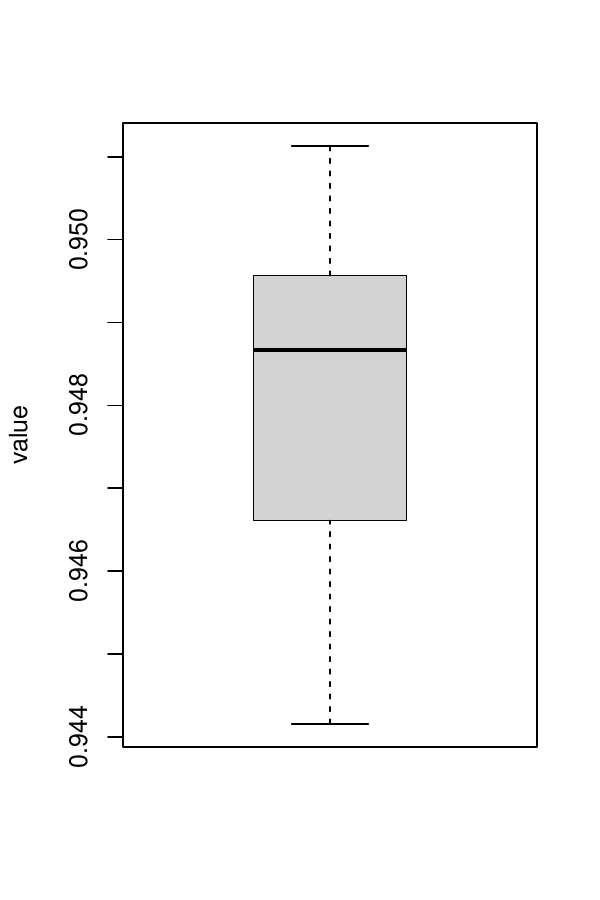}}\hfil
     \subfloat[Union of tree alone.]
  {\includegraphics[width=.22\textwidth]{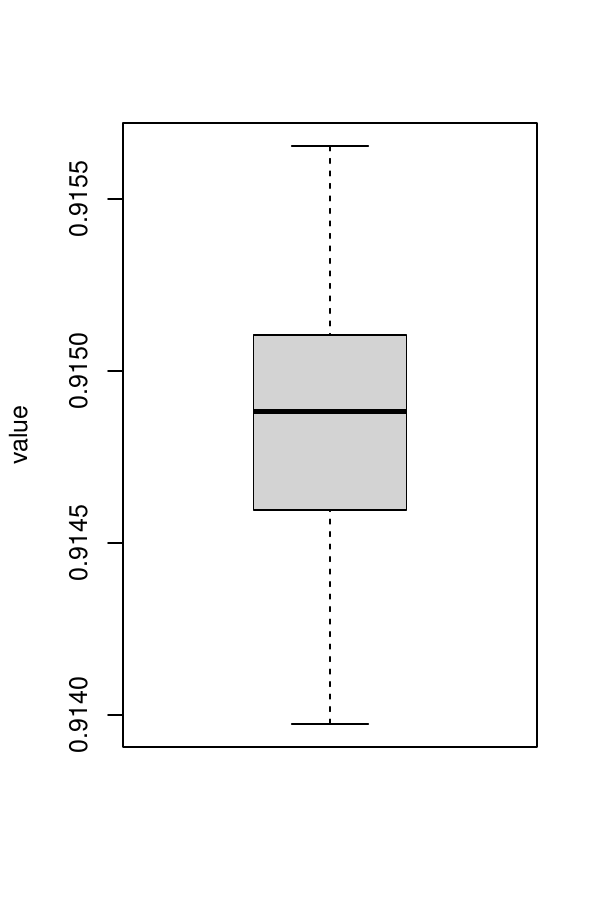}}\hfil
  \caption{Boxplot of the spectral norm of the companion-form matrices in the VAR($1$) equivalent representation for the VAR($d$) model, associated with each sampled $\bar C$ estimated in the data application.}
\end{figure}

\section{Additional Results from the Neuroimaging Data Analysis}

\begin{figure}[H]
  \centering
  \subfloat[Graph estimate using the lasso regularization. The graph has 1086 edges.]
	{\begin{minipage}{0.3\textwidth}
			  \includegraphics[width=1\textwidth]{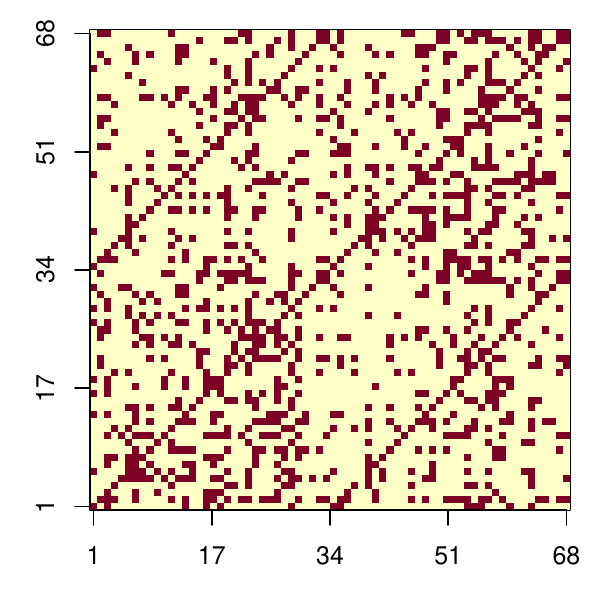}\\
  			\includegraphics[width=1\textwidth]{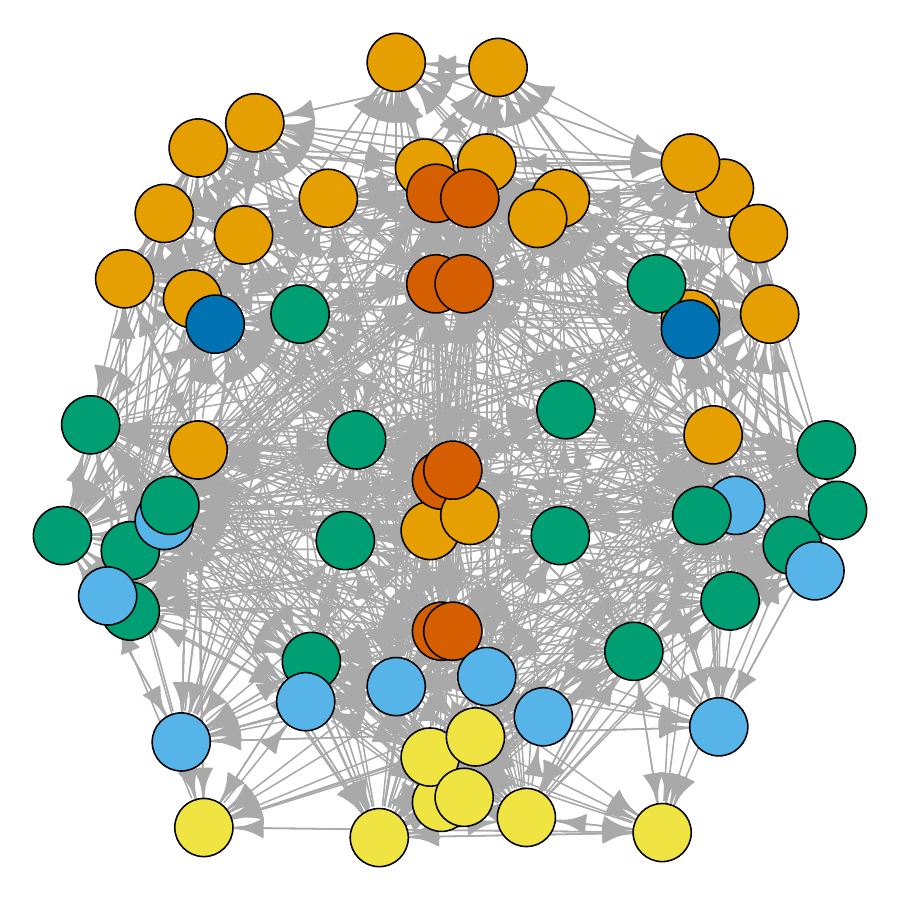}
	\end{minipage}
	}\;\;
  \subfloat[Graph estimate using the elastic net regularization. The graph has 1109 edges.]{\begin{minipage}{0.3\textwidth}
		  \includegraphics[width=1\textwidth]{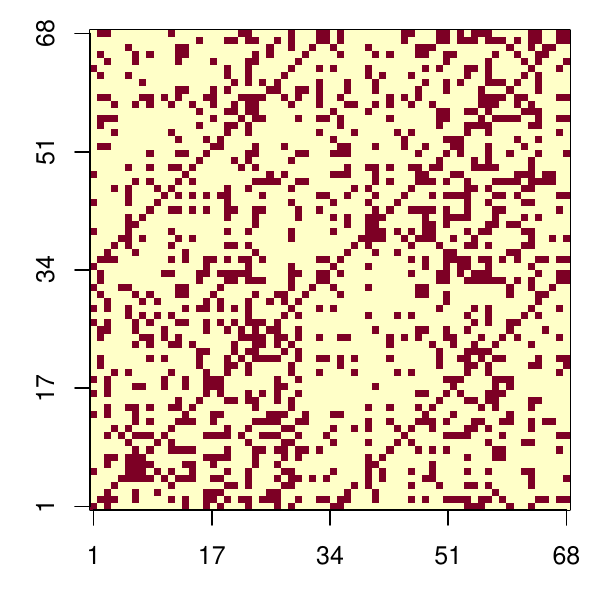}\\
  			\includegraphics[width=1\textwidth]{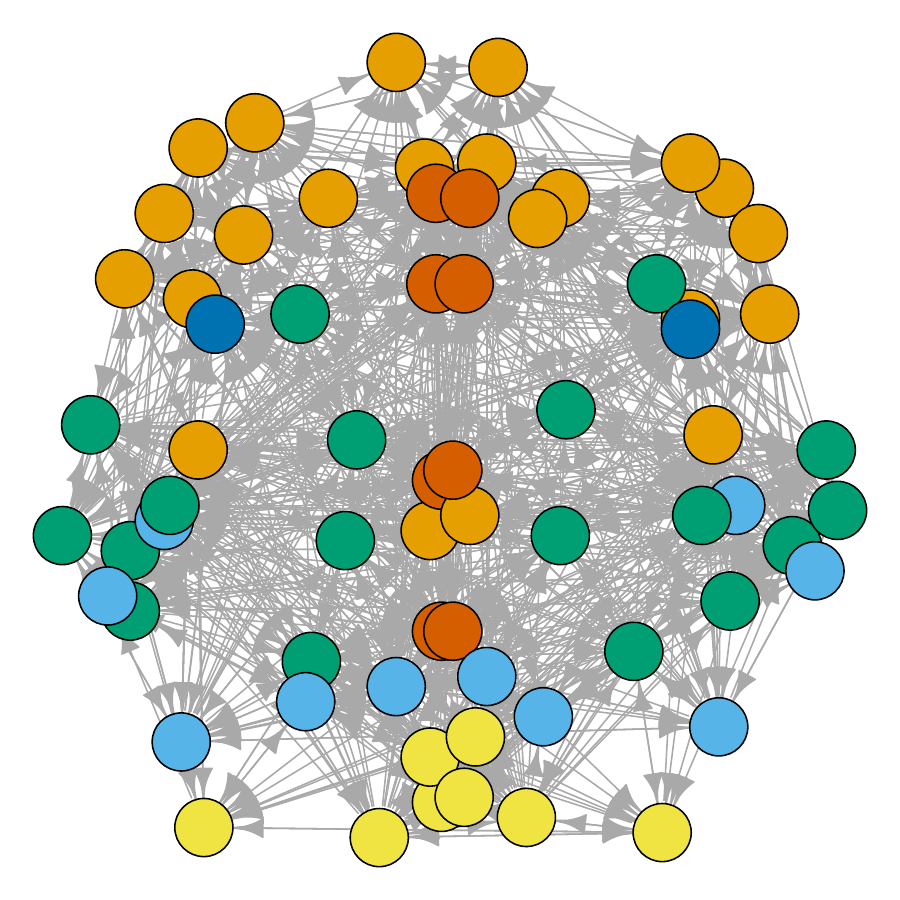}
	\end{minipage}
	}\;
  \caption{Comparing graph point estimates from several vector autoregressive models. Nodes are plotted using the  Desikan-Killiany atlas node coordinates and sized according to their degrees. Six cortical regions are shown in colors.}
  \label{fig:data_m_additional}
\end{figure}

\section{Additional Results on Area under the Curve Calculations}

\begin{figure}[H]
  \centering
             \subfloat[Area under the curve estimates in $G$ at $p=30$.]{\includegraphics[width=.45\textwidth]{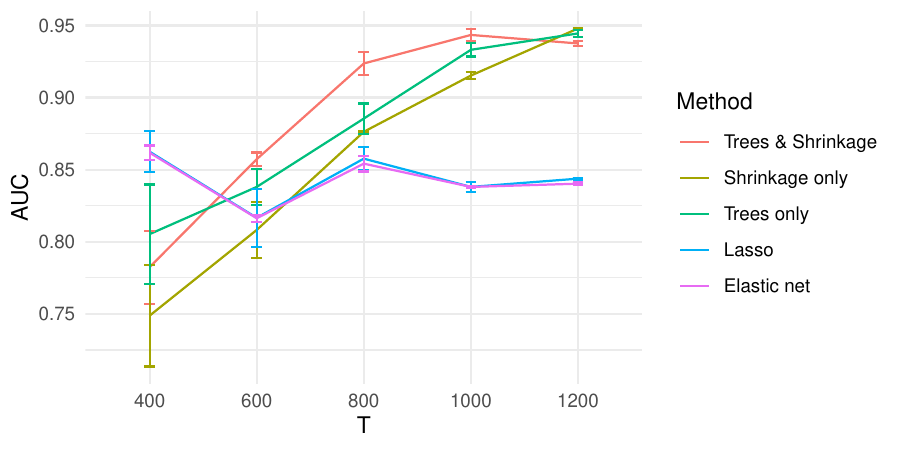}}\hfil
             \subfloat[Area under the curve estimates in $G$ at $p=80$.]{\includegraphics[width=.45\textwidth]{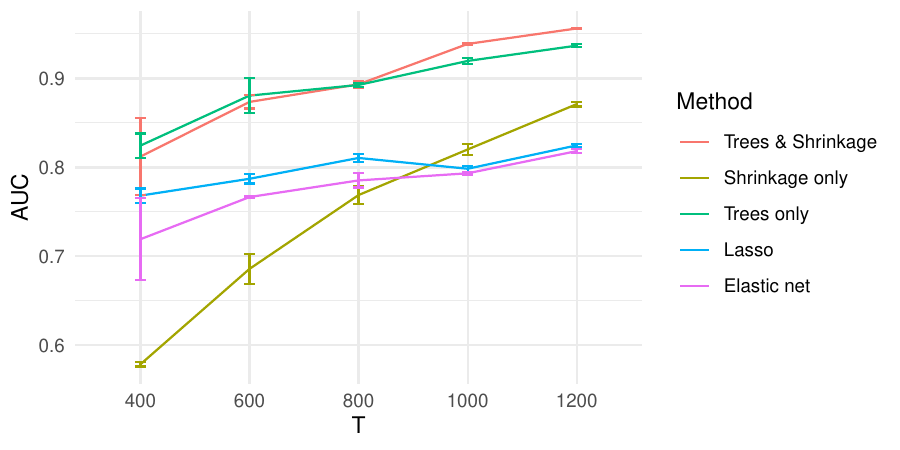}}\hfil
             \caption{Simulation results when the ground-truth graph $\bar G_0$ has a low tree-rank at $2$.  }
\end{figure}

\begin{figure}[H]
  \centering
             \subfloat[Area under the curve estimates in $G$ at $p=30$.]{\includegraphics[width=.45\textwidth]{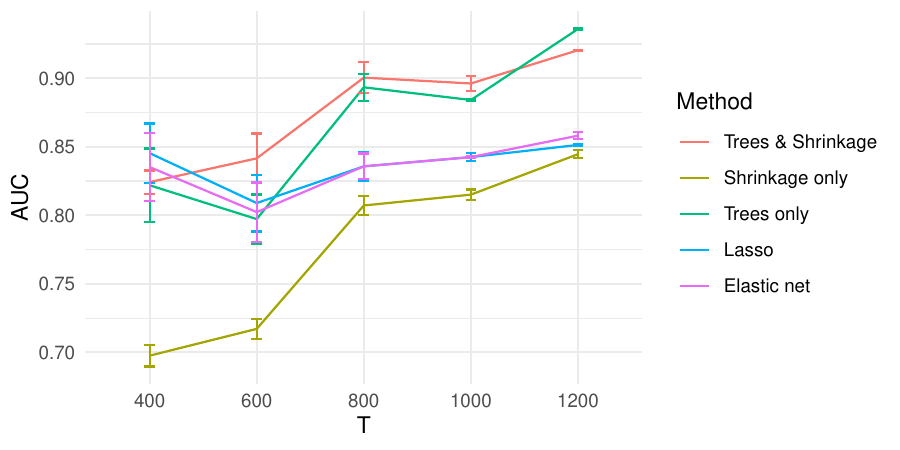}}\hfil
             \subfloat[Area under the curve estimates in $G$ at $p=80$.]{\includegraphics[width=.45\textwidth]{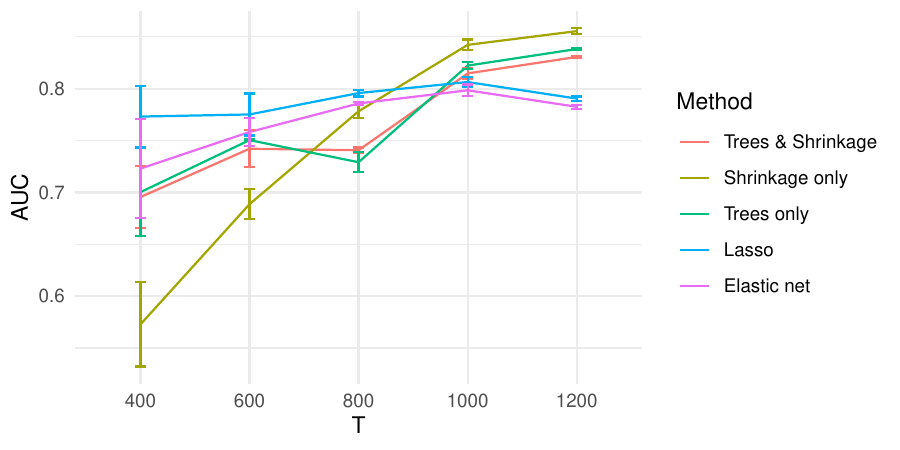}}\hfil
               \caption{Simulation results when the ground-truth graph $\bar G_0$ has a 95\% sparsity.}
\end{figure}

\section{Rapid Mixing of Markov Chains for the Gibbs Sampler}

\begin{figure}[H]
  \centering
             \subfloat[Trace and ACF plots for the regression coefficient $C_{1,1}^{(1)}$.]{\includegraphics[width=.45\textwidth, height=2.5cm]{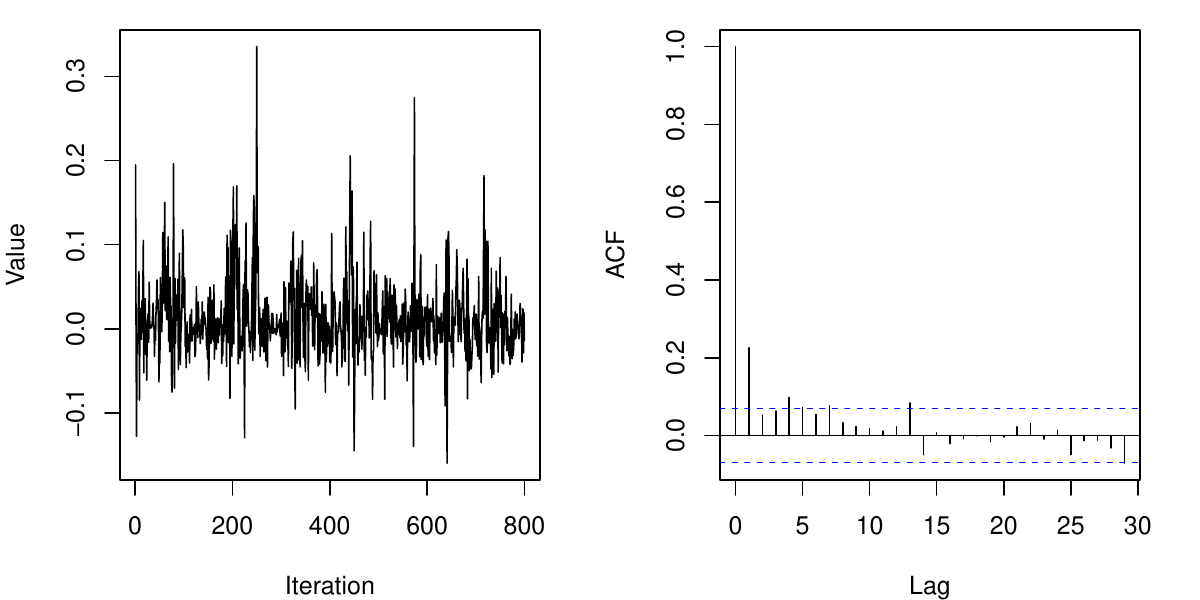}}\hfil
             \subfloat[Trace and ACF plots  for the regression coefficient $C_{10,3}^{(2)}$.]{\includegraphics[width=.45\textwidth, height=2.5cm]{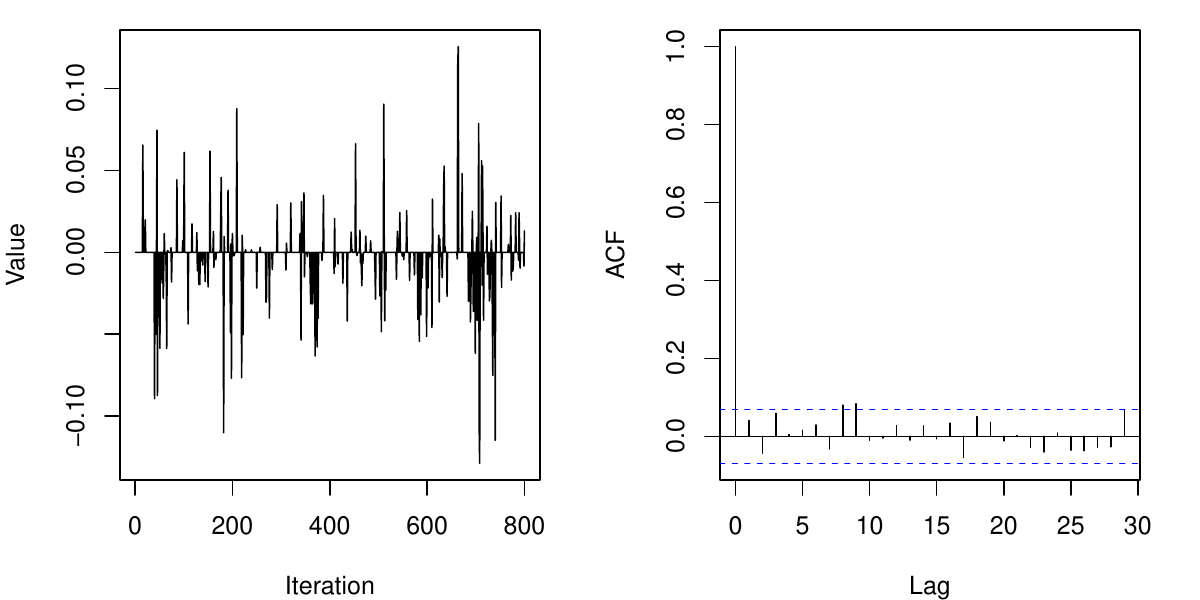}}\hfil
			 \subfloat[Trace and ACF plots  for the regression coefficient $C_{5,25}^{(3)}$.]{\includegraphics[width=.45\textwidth, height=2.5cm]{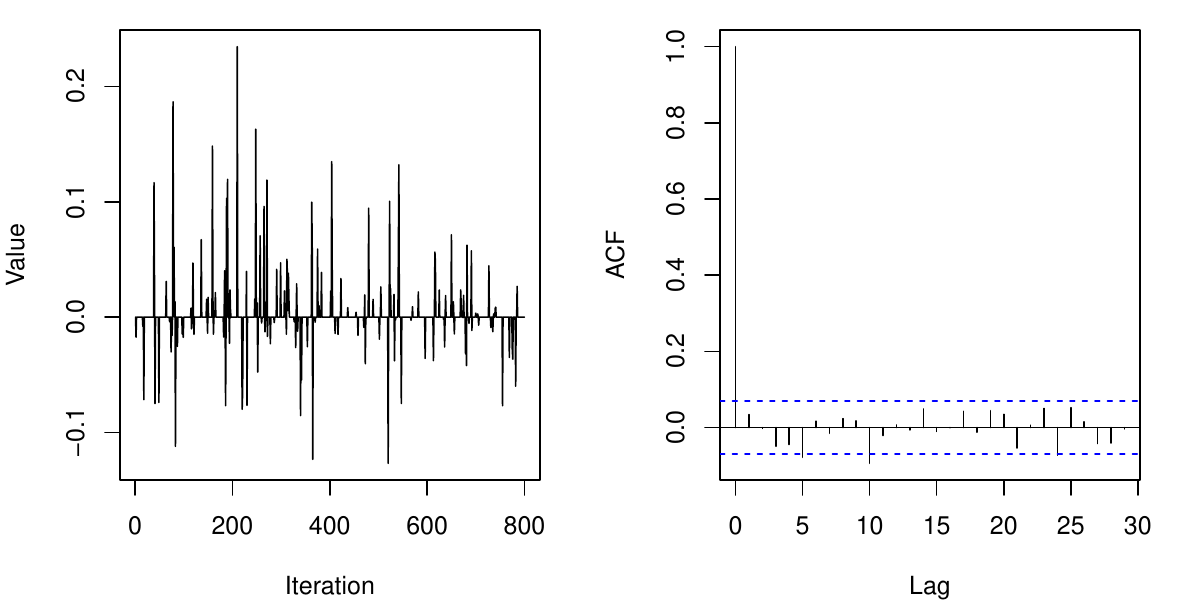}}\hfil
             \subfloat[Trace and ACF plots for the degree of node 1 in $A_T$.]{\includegraphics[width=.45\textwidth, height=2.5cm]{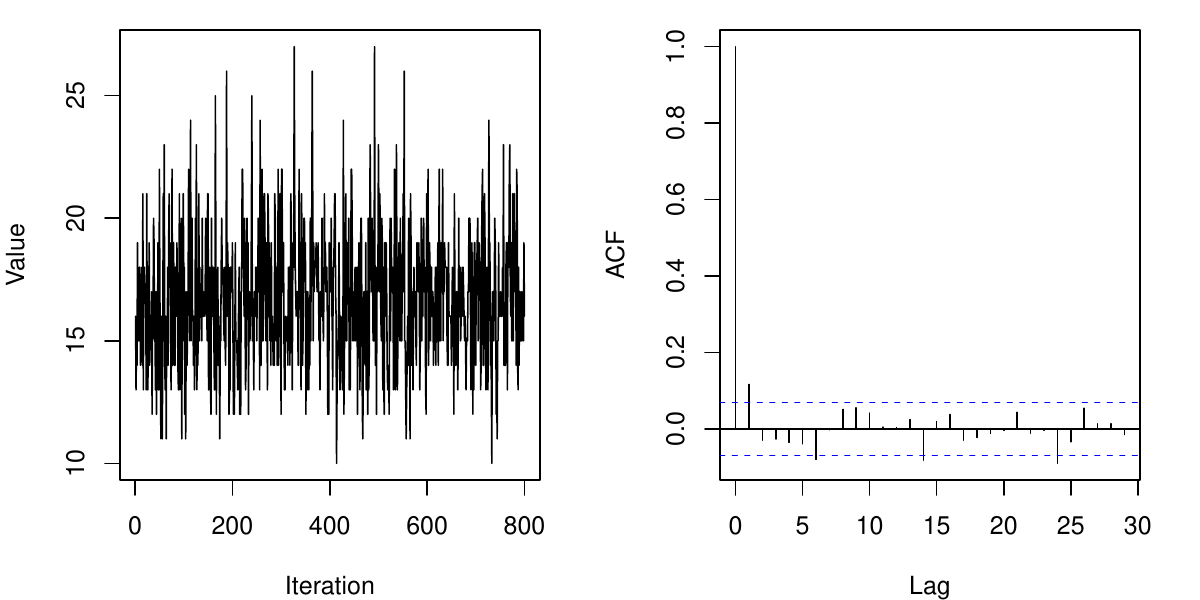}}\hfil
              \subfloat[Trace and ACF plots for the degree of node 40 in $A_T$.]{\includegraphics[width=.45\textwidth, height=2.5cm]{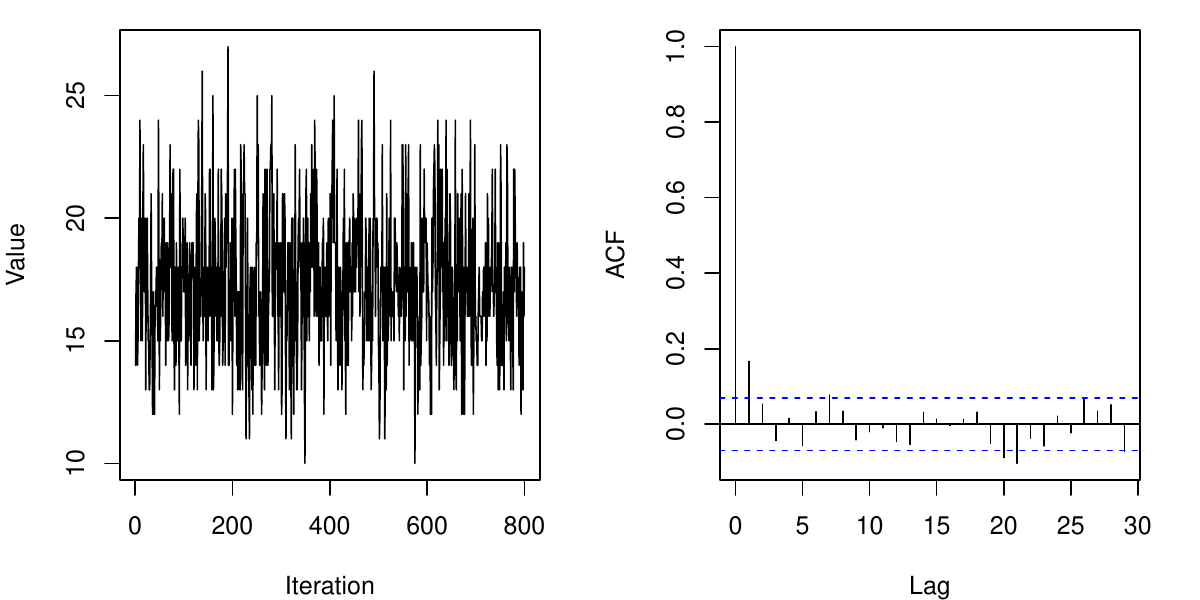}}\hfil
              \subfloat[Trace and ACF plots for the degree of node 80 in $A_T$.]{\includegraphics[width=.45\textwidth, height=2.5cm]{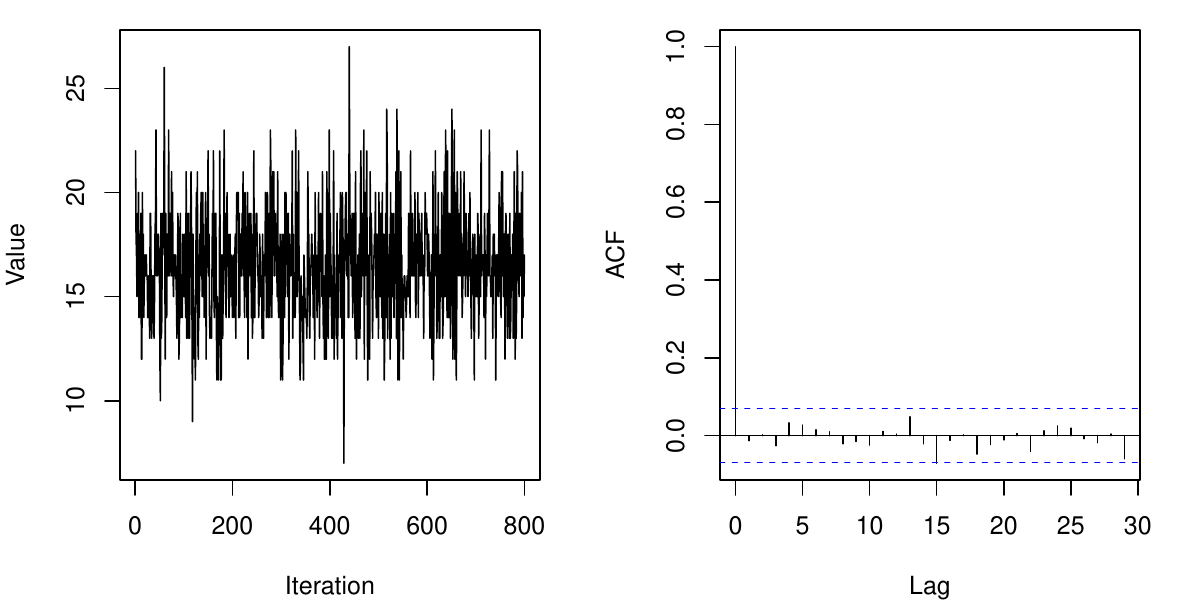}}\hfil
             \caption{Diagnostic plots showing rapid mixing of Markov chains for the Gibbs Sampler. The results are from the simulation with ground-truth graph $G_0$ having a low tree-rank at $2$, data generated at $p=80$, $\mathcal T=400$, and $d=3$. Panels (a-c) show the mixings in coefficient estimates (at three indices), and Panels(d-f) show the mixings in the union of trees estimate (using the degrees of three nodes).}
\end{figure}

\section{Additional Simulations on Sparse Graph Estimation}

\vspace{-1cm}
\begin{figure}[H]
  \centering
  \subfloat[Relative estimation error for $C$ at $p=30$.]{\includegraphics[width=.45\textwidth]{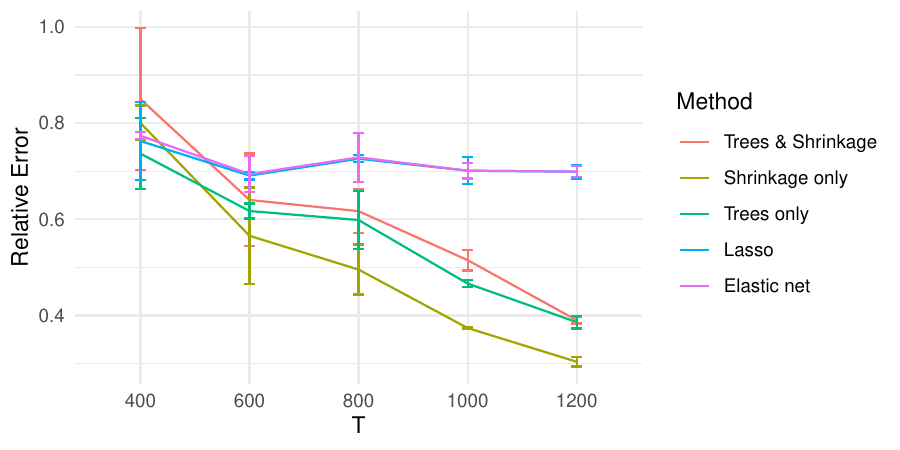}}\hfil
             \subfloat[Relative estimation error for $G$ at $p=30$.]{\includegraphics[width=.45\textwidth]{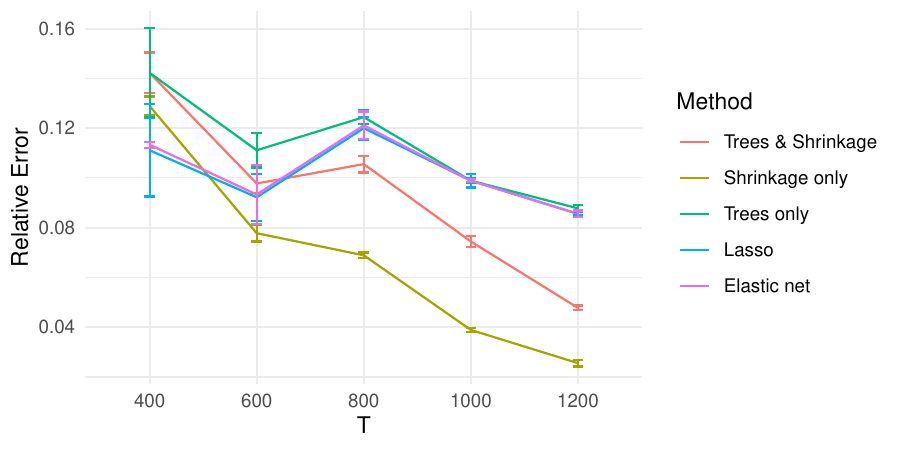}}\hfil
                          \subfloat[Area under the curve estimates in $G$ at $p=30$.]{\includegraphics[width=.45\textwidth]{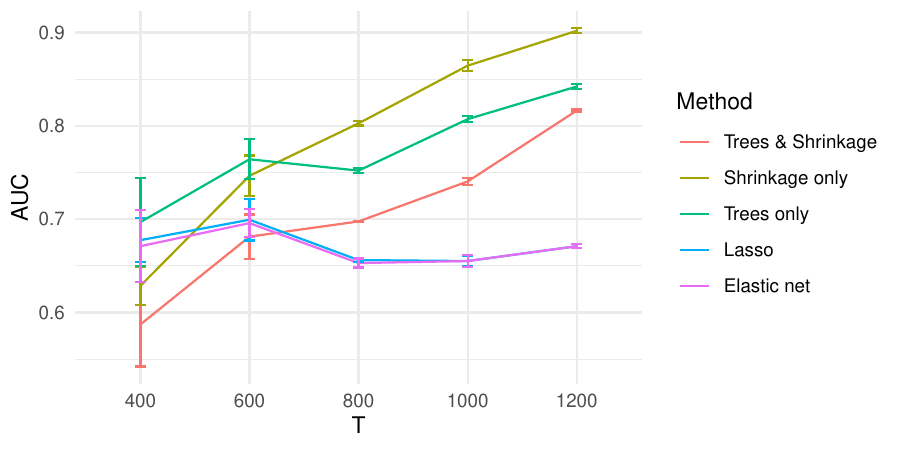}}\hfil
  \subfloat[Relative estimation error for $C$ at $p=80$.]{\includegraphics[width=.45\textwidth]{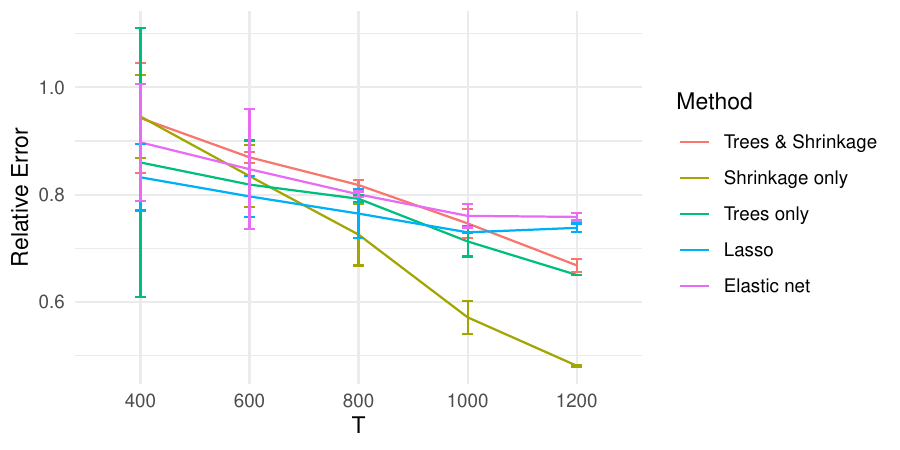}}\hfil
             \subfloat[Relative estimation error for $G$ at $p=80$.]{\includegraphics[width=.45\textwidth]{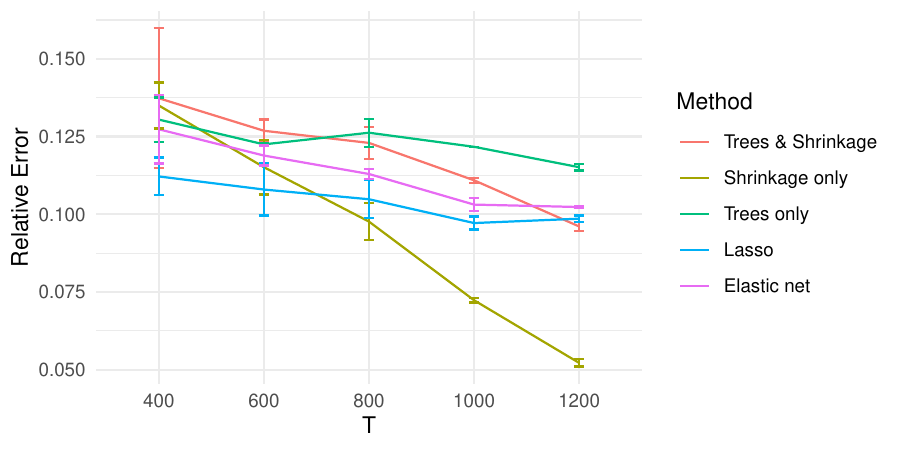}}\hfil
                          \subfloat[Area under the curve estimates in $G$ at $p=80$.]{\includegraphics[width=.45\textwidth]{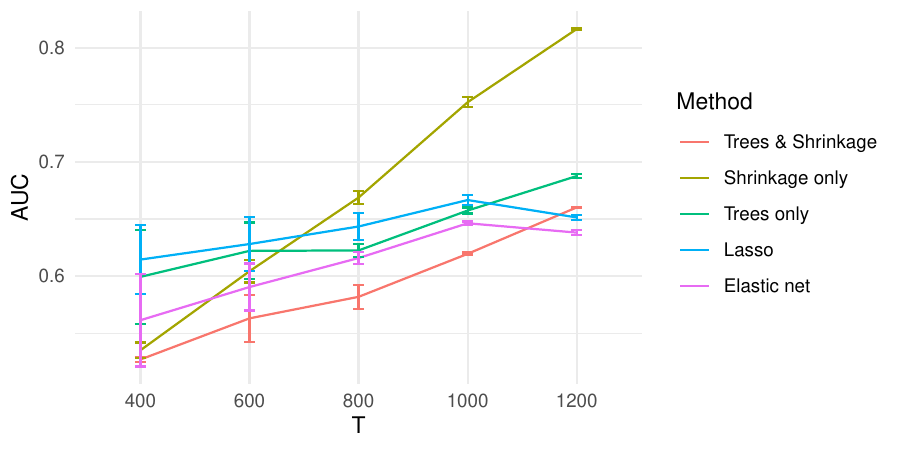}}\hfil               \caption{Simulation results when the ground-truth graph $G_0$ has a 87\%
               sparsity.  }
\end{figure}

\vspace{-1cm}
\begin{figure}[H]
  \centering
  \subfloat[Relative estimation error for $C$ at $p=30$.]{\includegraphics[width=.45\textwidth]{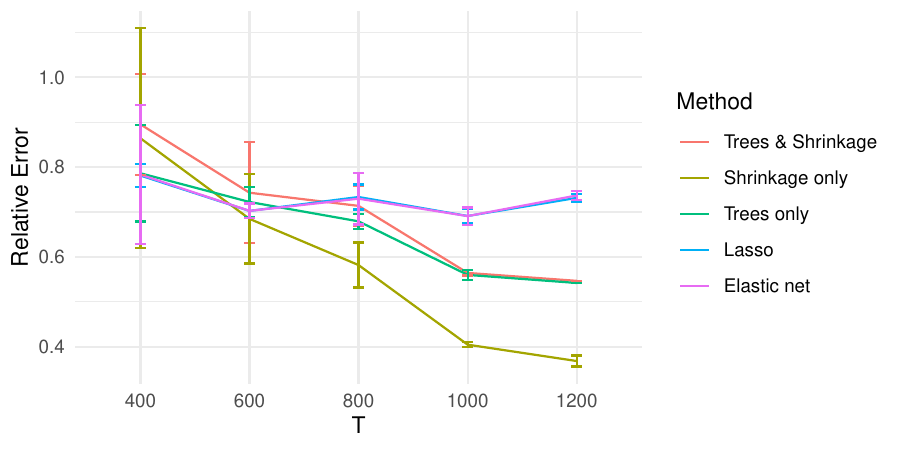}}\hfil
             \subfloat[Relative estimation error for $G$ at $p=30$.]{\includegraphics[width=.45\textwidth]{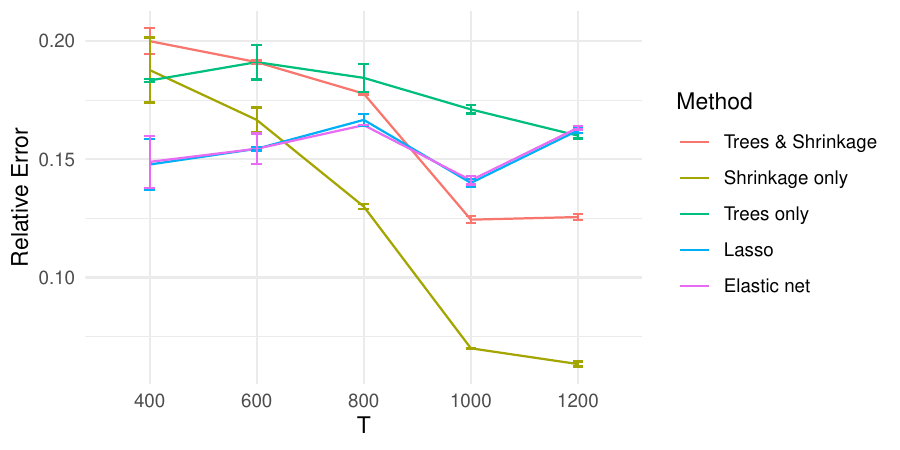}}\hfil
                          \subfloat[Area under the curve estimates in $G$ at $p=30$.]{\includegraphics[width=.45\textwidth]{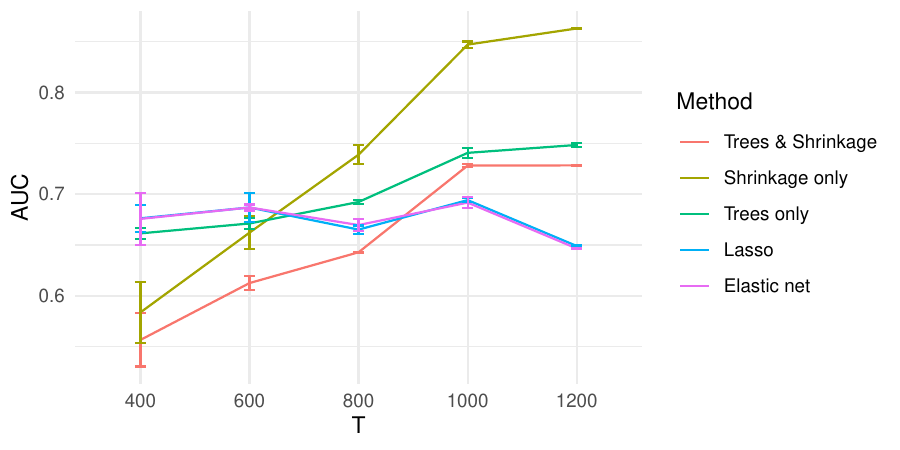}}\hfil
  \subfloat[Relative estimation error for $C$ at $p=80$.]{\includegraphics[width=.45\textwidth]{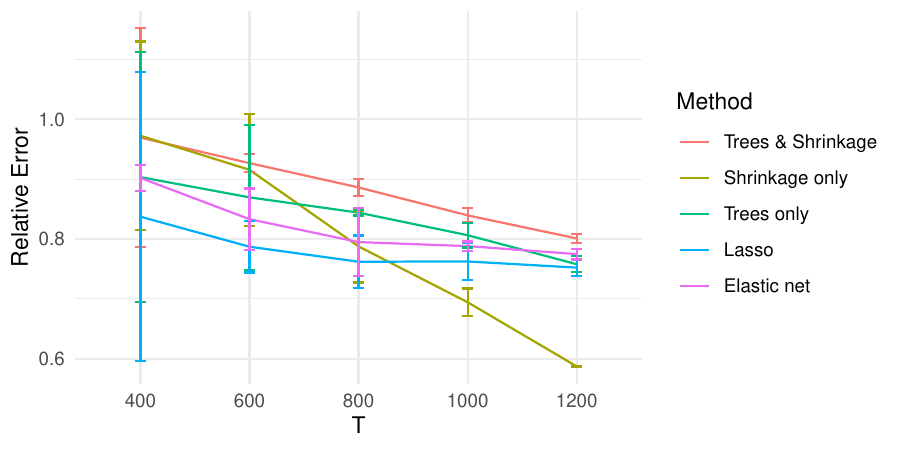}}\hfil
             \subfloat[Relative estimation error for $G$ at $p=80$.]{\includegraphics[width=.45\textwidth]{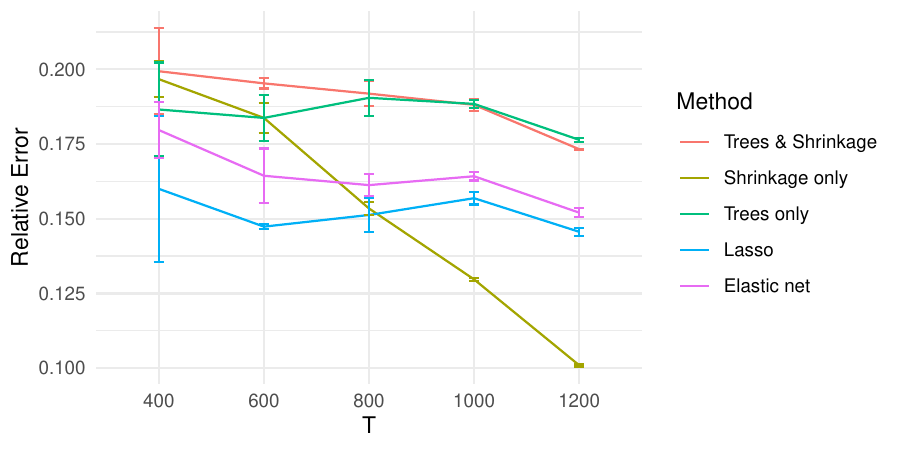}}\hfil
                          \subfloat[Area under the curve estimates in $G$ at $p=80$.]{\includegraphics[width=.45\textwidth]{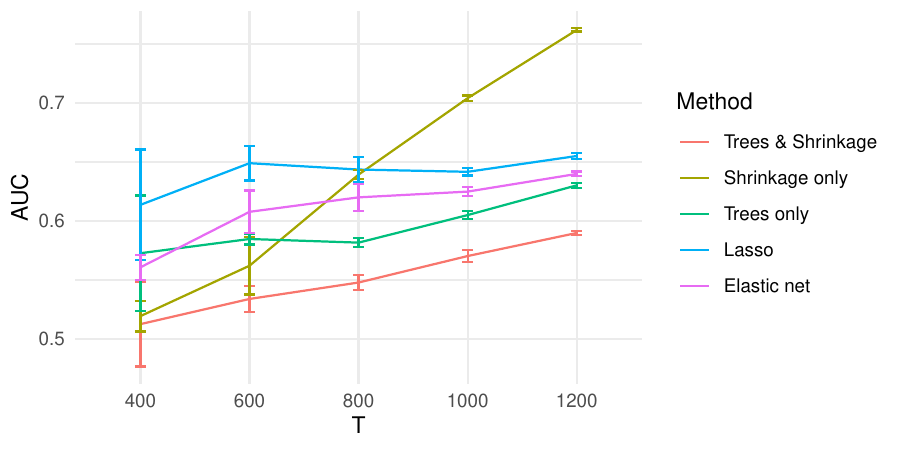}}\hfil               \caption{Simulation results when the ground-truth graph $G_0$ has a 80\%
               sparsity.  }
\end{figure}

\end{document}